\documentclass{article}
\usepackage[]{graphicx}
\usepackage{comment}
\usepackage[]{xcolor}
\usepackage{amsmath}
\makeatletter
\def\maxwidth{ %
  \ifdim\Gin@nat@width>\linewidth
    \linewidth
  \else
    \Gin@nat@width
  \fi
}
\makeatother

\definecolor{fgcolor}{rgb}{0.345, 0.345, 0.345}

\usepackage{booktabs}

\usepackage{framed}
\makeatletter
 {\par\unskip\endMakeFramed%
 \at@end@of@kframe}
\makeatother

\definecolor{shadecolor}{rgb}{.97, .97, .97}
\definecolor{messagecolor}{rgb}{0, 0, 0}
\definecolor{warningcolor}{rgb}{1, 0, 1}
\definecolor{errorcolor}{rgb}{1, 0, 0}
\newenvironment{knitrout}{}{} % an empty environment to be redefined in TeX

\usepackage[toc,page]{appendix}

\usepackage{alltt}
\usepackage[toc,page]{appendix}
\usepackage{enumerate} %for lists
\usepackage{graphicx}
\usepackage{amssymb}
\usepackage{amsthm}
\usepackage{setspace} % for spacing???
\usepackage{xcolor}
\usepackage[utf8]{inputenc}
\usepackage[english]{babel}
\usepackage{natbib} 
\usepackage{caption}
\usepackage{subcaption} %subcaption stuff for diagrams with parts (a),(b),etc.
\usepackage[version=3]{mhchem} %chemistry stuff
\usepackage{graphicx}    % Any additional package
\usepackage[colorlinks=true, urlcolor=blue]{hyperref} % Load hyperref last

\author{Chunlei Ge \&
        W. John Braun \\UBC Okanagan}
\title{Differential Equation-Constrained Local Regression for Data with Sparse Design}

\begin{document}
\maketitle

%% for pretty printing and a nice hypersummary also set:
%\Plainauthor{Chunlei Ge, W. John Braun} %% comma-separated
%\Plaintitle{Quick and Simple Kernel Differential Equation Regression Estimators} %% without formatting
%\Shorttitle{Differential Equation Regression} %% a short title (if necessary)

%% an abstract and keywords
\begin{abstract}
 Local polynomial regression of order one or higher often performs poorly in areas with sparse data. In contrast, local constant regression tends to be more robust in these regions, although it is generally the least accurate approach, especially near the boundaries of the data.  Incorporating information from differential equations which may approximately or exactly hold is one
 way of extending the sparse design capacity of local constant regression while reducing bias and variance.   A nonparametric regression method that exploits first order differential equations is studied in this paper and applied to noisy mouse tumour growth data. Asymptotic biases and variances of kernel estimators using Taylor polynomials with different degrees are discussed. Model comparison is performed for different estimators through simulation studies under various scenarios which simulate exponential-type growth. 
\end{abstract}


%% publication information
%% NOTE: Typically, this can be left commented and will be filled out by the technical editor
%% \Volume{13}
%% \Issue{9}
%% \Month{September}
%% \Year{2004}
%% \Submitdate{2004-09-29}
%% \Acceptdate{2004-09-29}

%% The address of (at least) one author should be given
%% in the following format:

\noindent
\textbf{Address.}

\noindent
  Chunlei Ge\\
  Department of Computer Science, Mathematics, Physics and Statistics, UBCO\\
  Kelowna, BC V1V 1V7\\
  E-mail: \texttt{chunlei.ge@ubc.ca}\\
  URL: \url{https://cmps.ok.ubc.ca}

%% It is also possible to add a telephone and fax number
%% before the e-mail in the following format:
%% Telephone: +43/1/31336-5053
%% Fax: +43/1/31336-734

%% for those who use Sweave please include the following line (with % symbols):
%% need no \usepackage{Sweave.sty}
%% end of declarations %%%%%%%%%%%%%%%%%%%%%%%%%%%%%%%%%%%%%%%%%%%%%%%

%% include your article here, just as usual
%% Note that you should use the \pkg{}, \proglang{} and \code{} commands.

\newcommand{\Sconcordance}[1]{%
  \ifx\pdfoutput\undefined%
  \csname newcount\endcsname\pdfoutput\fi%
  \ifcase\pdfoutput\special{#1}%
  \else%
   \begingroup%
     \pdfcompresslevel=0%
     \immediate\pdfobj stream{#1}%
     \pdfcatalog{/SweaveConcordance \the\pdflastobj\space 0 R}%
   \endgroup%
  \fi}
\Sconcordance{concordance:CurrentVersion.tex:CurrentVersion.Rnw:%
1 74 1 50 0 1 23 7 1 1 2 3 1 10 0 1 1 1 3 45 1 1 6 5 1 13 0 7 1 8 0 33 %
1 1 21 21 1 1 9 171 1 1 14 249 1 1 25 26 1 1 19 60 1 1 3 11 1 10 0 39 1}

%% for those who use Sweave please include the following line (with % symbols):
%% need no \usepackage{Sweave.sty}
%% end of declarations %%%%%%%%%%%%%%%%%%%%%%%%%%%%%%%%%%%%%%%%%%%%%%%

%% include your article here, just as usual
%% Note that you should use the \pkg{}, \proglang{} and \code{} commands.

 \section{Introduction}

Modelling data involving relationships between outcome variables and potential explanatory variables remains an active area of research. Generalized linear models, first described by \citet{mccullagh1989}, are a popular class of parametric regression models because they provide clear and interpretable insights into how predictors influence outcomes. However, compared to nonparametric regression, parametric approaches are more interpretable but less flexible and accurate due to constraints imposed on the functional relationships between response and explanatory variables, which must be specified in advance. In practical scenarios, scatterplots that do not fit well with straight lines or predefined parametric curves can often be better analyzed using nonparametric methods. These approaches adapt to the underlying functional relationship and reveal unexpected features in the data.

\citet{nadaraya1964estimating} and \citet{watson1964smooth} introduced the kernel regression estimator, now known as the Nadaraya–Watson or local constant regression estimator—a special case of the larger class of local polynomial regression estimators.  These estimators smooth scatter plots by averaging nearby values weighted by a kernel function indexed by a smoothing parameter, or bandwidth,  $h$. The books by \citet{fan1996local} and \citet{loader1999local} offer comprehensive methods for bandwidth selection. 

Local polynomial regression offers several benefits, including adaptive smoothing, local interpretability, and the absence of a need for explicit functional specifications. However, the reliance on local neighborhoods presents challenges in regions with sparse data, because of the number of local parameters that need to be estimated.  Incorporating information from differential equations can provide valuable assistance in overcoming these limitations. The paper by \citet{ding2014estimation} focuses on parameter estimation in differential equations but also introduces differential equation-constrained local polynomial regression.   We note also the similar one-step parameter estimation strategy for  differential equations proposed by \citet{hall2014quick}.  
Alternatively, \citet{LiangWu2008} propose a kernel-based differential equation (DE) estimation method using a measurement error model, which, while related, differs from the approach presented in this paper.  Their work also provides a comprehensive review of the literature on fitting ordinary DEs to data.  

  DE-constrained estimation also aligns with recent research efforts incorporating additional information as estimation constraints, as demonstrated by \citet{chatterjee2016constrained} and \citet{dai2024kernel}.  The paper of \citet{shan2023low} considers Principal Component Analysis using kernel methods and DEs.  Data sharpening subject to constraints, introduced by \citet{braun2001data}, is a technique in which data values are adjusted to improve statistical estimators while satisfying specific constraints.
An important reference in the area of DE parameter estimation and fitting is \citet{ramsay2007parameter}, which introduces a spline-based approach requiring extensive tuning.  \citet{carey2021fast} use spline methods to estimate unknown parameters in linear dynamical systems. 

The DE-constrained local polynomial regression approach studied in this paper effectively reduces asymptotic bias without substantially increasing variance. 
%This approach represents a specific form of differential equation-constrained local polynomial regression. 
By incorporating DEs, local polynomial regression becomes capable of handling more complex data types, such as those generated from sparse designs.  The approach is applicable to a broad class of DEs, including both linear and nonlinear forms. In this paper, we focus on a relatively simple model: the univariate DE-constrained local exponential growth model.  This provides, essentially, a case study, which transparently demonstrates the potential of the method.   In particular, we will 
highlight how the DE-constrained method performs on sparsely designed
data. 
 
\subsection{Differential Equation-Constrained Regression Model}

Our main focus is on models constrained by first-order DEs.
Given $n$ independent observations on an explanatory variable  $x_0$ and
a response variable $y$,   
we consider models of the form 
%\[y_i = g(x_i) + \varepsilon_i \mbox{ where } g'(x_0) = F(x_0,g(x_0)), \ \ \ \ i = 1, 2, \ldots, n,  \]

\begin{equation}
y_i = g(x_i) + \varepsilon_i \mbox{ where } g'(x_0) = F(x_0,g(x_0)), \ \ \ \  x_i \in [a,b], \ \ i = 1, 2, \ldots, n.  
\label{equ:general} 
\end{equation}
for some Lipschitz continuous function $F$ and uncorrelated, mean-zero errors $\varepsilon_i$.   
We assume that the design points are randomly sampled from an interval $[a, b]$
according to a probability density function $f$, or have been selected according to
a fixed sampling design within that interval.  

\subsection{Exponential Growth Model}

As a case study to demonstrate our approach, we perform the 
regression analysis of the above model using the specific form, $F(x_0,g(x_0))=\lambda g(x_0)$. In this case, the model is the exponential growth model:
\begin{equation}
y_i = g(x_i) + \varepsilon_i \mbox{ where } g'(x_0) = \lambda g(x_0), \ \ \ \  x_i \in [a,b], \ \ i = 1, 2, \ldots, n.  
\label{equ:exp} 
\end{equation}

 The explicit solution to this DE is 
\begin{equation}\label{expnls} g(x) = g(a) e^{\lambda (x - a)}.  \end{equation}
This model can be directly fit to the data by estimating $\lambda$ and the initial condition, $g(a)$,  through nonlinear least-squares regression
or by linear regression applied to the log-linear version of the model.   The choice of approach depends upon whether the error structure is
additive or multiplicative.  

\subsection{Sparse Tumor Growth Data }
In what follows, we will consider a set of control data from a chemotherapy trial in an animal experiment. The mouse tumor data (\citet{plume1993relative}) were collected on mice tumor volumes over time. Tumor volume measurements in Table \ref{table:fullmouse} were taken from a single mouse. Times are recorded in days, and volumes are in cubic centimeters.  

 \begin{table}[ht]
\centering
\scalebox{1}{
\begin{tabular}{ccc}
  \hline
  & time & volume \\ 
  \hline
1  &  21  & 0.05 \\
2  &  25  & 0.09 \\
3  &  28  & 0.22 \\
4  &  31  & 0.32 \\
5  &  33  & 0.61 \\
6  &  35  & 0.70 \\
7  &  38  & 0.90 \\
8  &  40  & 1.29 \\
9  &  42  & 1.77 \\
10 &  45 &  3.32 \\
  \hline  
\end{tabular}
}
\caption{The full set of observations on mouse tumor volume.} 
\label{table:fullmouse}
\end{table}

\subsection{Applying Local Regression to Data Subject to Sparse Design}

In order to demonstrate the effectiveness of the DE-constrained approach, 
we  simulated a larger dataset (sample size = 41) based on a fitted nonlinear regression model (\ref{expnls})
applied to the above mouse tumor data.

To illustrate the performance of various local polynomial estimators on sparse data, we artificially removed some data points from the simulated mouse tumour data.  The resulting data set is displayed as filled black circles in Figure \ref{fig:sparse}, while the deleted data are represented by open red circles.  
%Although the sparse dataset contains only five data points and is generally unsuitable for local polynomial regression models applied, it will provide a useful example to demonstrate the usefulness of the constrained-DE method.  We limit our analysis to local constant, local linear, and local quadratic regression. The local cubic regression model is excluded, as it requires estimating four parameters, and the dataset is not large enough to support it. 

The plot in the right panel of Figure \ref{fig:sparse} illustrates the results of this approach for a grid of evaluation points $x_0$ within the sparse interval. The bandwidth was selected as half of the median of the successive differences in the sorted covariate values.  
 Specifically, we fit local constant, local linear, and local quadratic models to the second dataset using a Gaussian kernel with a bandwidth of $h = 3.5$ for each method. 
 As shown in the first three plots, each approach exhibits deficiencies due to the gap in the data. The local linear and quadratic models introduce spurious bumps, while the local constant model suffers from severe inaccuracies.
Local constant regression can sometimes handle sparse data, but often lacks accuracy, particularly near boundaries.  Higher-order local polynomial methods tend to fail in these situations. 
By contrast, the DE-constrained method, which is the subject of study in this paper, was applied and plotted in the right-most panel.  
Compared to the other curves in the figure, the DE seems to have provided valuable information.  

\begin{knitrout}\scriptsize
\definecolor{shadecolor}{rgb}{0.969, 0.969, 0.969}\color{fgcolor}\begin{figure}
{\centering \includegraphics[width=1\textwidth]{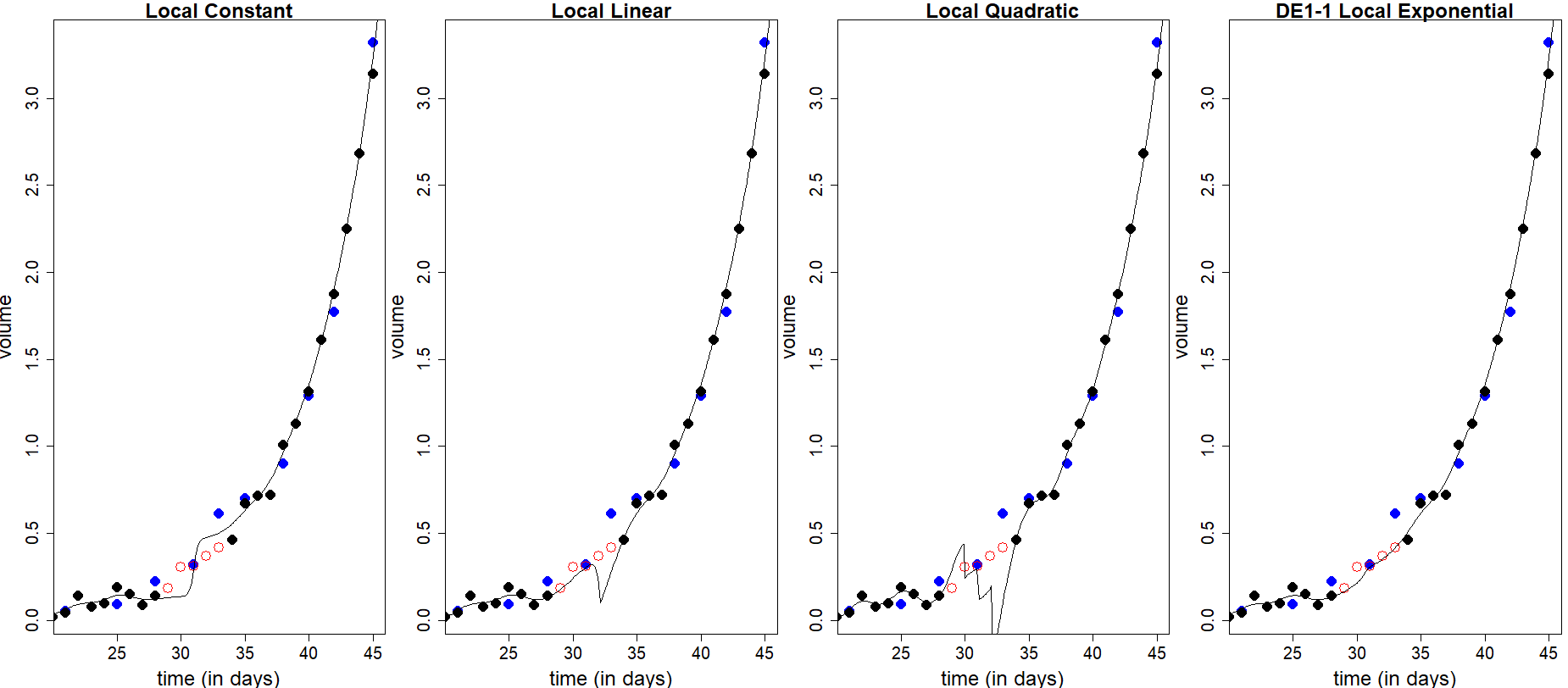} 
}
\caption{Simulated mouse tumour growth data with artificially removed observations (shown as open red circles). The original mouse tumor growth data shown as filled blue circles.  Left panel: Local constant model;  Center-left panel: Local linear model;   Centre-right panel: Local quadratic model; Right panel: Local growth model.  All models are fit to the sparse data (filled black circles) using a Gaussian kernel with bandwidth $h = 3.5$.}\label{fig:sparse}
\end{figure}

\end{knitrout}

\subsection{The Outline of the Paper}
  
The remainder of this paper is organized as follows.  Section 2 outlines the DE-constrained Local Polynomial Regression method. 
Our objective is to leverage the low variance and data sparseness-handling capabilities of local regression estimators which rely on a minimal number of local parameters (ideally, one), while attaining the accuracy of higher-order local polynomial methods. To achieve this, we assume that the regression function is the solution of a DE. 
The theoretical demonstration for the local exponential growth model, which used first-degree DE-constrained approach, is presented in Section 3, showing that the bias is reduced compared to local constant regression without significantly increasing variance. In Section 4, the theory and method for bandwidth selection in large sample sizes are discussed. Section 5 features a simulation study comparing the proposed method to conventional local polynomial approaches and nonlinear least-squares estimation of the DE solution, while also exploring its effectiveness for sparse designs. The final section concludes with a discussion, touching on robustness of the method to model misspecification, as well as suggestions for future work.

\section{DE-Constrained Local Polynomial Regression}

The approach we describe here adapts local polynomial regression as described by  \citet{fan1996local} and extends the scope of the models beyond those considered by \citet{ding2014estimation}.  
To this end, we assume the design density, kernel, and bandwidth satisfy the conditions outlined in \citet{fan1996local}.
Specifically, the kernel function $K_h(x_0)$, which controls the weights of the observations, is a symmetric probability density function scaled by a smoothing parameter or bandwidth $h$. The kernel function $K_h(u)$ satisfies the regularity conditions of normalization and has a finite second moment.
\begin{comment}
\begin{itemize}
\item Symmetry:
\[K(u)=K(-u),\]
\item Non-negativity:
$$ K(u) \geq 0 \quad \text{for all} \quad u, $$
\item Normalization:
$$\int_{-\infty}^{\infty}K(u)du=1,$$
\item Finite second moment:
$$\int_{-\infty}^{\infty}u^2K(u)du<\infty,$$
\end{itemize}
\end{comment}
Additionally, $K_h(u)$ is sufficiently smooth to allow differentiation, enabling analysis of the bias and variance of the estimator.
The bandwidth $h$ satisfies the standard conditions: $n\rightarrow \infty,$ $h\rightarrow 0,  \text{ and } nh \rightarrow \infty.$

For a given evaluation point $x_0$, a DE-constrained local polynomial regression estimator for $g(x_0)$ is obtained
by minimizing the local least-squares objective function:
\begin{equation}\label{eqn:genmethod} \sum_{i=1}^n \{y_i - g_k^*(x_i)\}^2 K_h(x_i-x_0). \end{equation}
where 
\begin{equation}\label{eqn:star} g_k^*(x_i) = g(x_0) + (x_i-x_0) g'(x_0) + \cdots + \frac{(x_i-x_0)^k}{k!} g^{(k)}(x_0)  \end{equation}
is the $k$th degree Taylor approximation of $g(x_i)$ about $x_0$.  
By incorporating information from the DE, we have
\[ g'(x_0) = F(x_0, g(x_0)) \]
so that
\[ g_1^*(x_i) = g(x_0) + (x_i - x_0)F(x_0, g(x_0)). \]
Additional derivatives can be taken.  For example,  
\[ g''(x_0) = F_1(x_0, g(x_0)) + F_2(x_0, g(x_0)) g'(x_0) = F_1(x_0, g(x_0)) + F_2(x_0, g(x_0))F(x_0, g(x_0)), \]
where $F_j$ denotes the partial derivative of $F$ with respect to its $j$-th argument.   Thus,
\[ g_2^*(x_i) = g(x_0) + (x_i - x_0)F(x_0, g(x_0)) + \frac{(x_i-x_0)^2}{2}\{F_1(x_0, g(x_0)) + F_2(x_0, g(x_0))F(x_0, g(x_0))\}. \]
Only one local parameter, $\alpha = g(x_0)$, is unknown and needs to be estimated.  
Global parameters can be estimated using the methods of \citet{ding2014estimation} or \citet{hall2014quick}, for example.

 We note, in passing, that 
higher order DE information might be available instead.  For example, if we have a second-order differential
equation of the form
\[ g''(x_0) = F(x_0, g(x_0), g'(x_0)) \]
we would write
\[ g_2^*(x_i) = g(x_0) + (x_i - x_0)g'(x_0) + \frac{(x_i-x_0)^2}{2}F(x_0, g(x_0), g'(x_0)). \]
There are two local parameters to be estimated: $\alpha = g(x_0)$ and $\beta = g'(x_0)$.    Observe
that $g_1^*(x_i) = g(x_0) + (x_i - x_0)g'(x_0)$ which corresponds to the local linear regression model.  

In general, the local parameters would be estimated using a nonlinear optimizer.  Linear methods are applicable 
for data governed by linear differential equations.  The first order linear case yields the statistical model 
\begin{equation}
y_i = g(x_i) + \varepsilon_i \mbox{ where } g'(x_0) = a(x_0)g(x_0)+b(x_0), \ \ \ \  x_i \in [a,b], \ \ i = 1, 2, \ldots, n.  
\label{eqn:generallinear} 
\end{equation}
where $a(x_0)$ and $b(x_0)$ are known functions possessing $k+2$ continuous derivatives.
In this case, the first two local approximants to $g(x_0)$  are
\[ g_1^*(x_i) = g(x_0) + (x_i - x_0)\{a(x_0)g(x_0)+b(x_0)\} \]
and
\[ g_2^*(x_i) = g(x_0) + (x_i - x_0)g'(x_0) + \frac{(x_i-x_0)^2}{2}\{a'(x_0) a(x_0) g(x_0) + a'(x_0) b(x_0) + b'(x_0)\}. \]
Higher degree approximants can be computed easily using the recursive relationship
 \[ g^{(k+1)}(x_0) =\sum_{l=0}^{k}\binom{k}{l}a^{l}(x_0)g^{(k-l)}(x_0)+b^{(k)}(x_0). \]

The $k$th-degree version of the estimation technique amounts to minimizing (\ref{eqn:genmethod}) with respect to the unknown local parameter(s).  
For example,  the first degree estimator, using first order DE information, is obtained by minimizing 
\[  \sum_{i=1}^n \{ y_i - \alpha -  (x_i - x_0)F(x_0, \alpha)\}^2 K_h(x_i-x_0)  \]
with respect to $\alpha$.  We refer to the resulting estimator  $\widehat{\alpha} = \widehat{g}(x_0)$ as the local DE1-1 estimator,
where the first `1' refers to the order of the DE and the second `1' refers to the degree of the Taylor expansion
employed.    

\section{Local Exponential Growth Model}
In this paper, we have chosen to work through all theoretical details in a simple special case in order for the benefits (and drawbacks) of the method to be as transparent as possible.  Furthermore, the specific set of estimators we obtain here will have wide application, because of the ubiquity of exponential-growth type data that arise in a large variety of scientific areas.  
Specifically, we assume response values that follow the exponential growth model (\ref{equ:exp}). 
This is a special case of (\ref{eqn:generallinear}) with $a(x_0)=\lambda$ and $b(x_0)=0$.

A local solution to the DE can be obtained by constructing a Taylor expansion of $g(x_i)$ around an evaluation point $x_0$, as
described in the previous section.
Our motivation for combining the differential equation with local regression to develop a differential equation-constrained estimation approach lies in its advantages: the new method is not constrained by the requirement for an initial value of the differential equation, and it remains robust to some model misspecifications, which we will discuss in another paper.  
We refer to this model as the local exponential growth model. 

Applying the first-degree Taylor expansion for $g(x_i)$ in a sufficiently small neighborhood of the evaluation point  $x_0$, where $x \in (a,b)$, we obtain:
\[ 
\sum_{i=1}^n \{y_i - g_1^*(x_i)\}^2 K_h(x_i - x) 
= \sum_{i=1}^n \{ y_i - g(x_0) - \lambda g(x_0) (x_i - x)\}^2 K_h(x_i-x_0).  \]
Optimizing the expression on the right leads to the DE-constrained local linear regression estimator DE1-1
\[ \widehat{g}_1(x_0) = \frac{
\sum_{i=1}^n y_i(1+(x_i-x_0)\lambda)K_h(x_i-x_0)}{\sum_{i=1}^n (1 + (x_i-x_0)\lambda)^2 K_h(x_i-x_0)}. \]
Similarly, we consider the DE-constrained local quadratic regression estimator. Applying the second-degree Taylor expansion for $g(x_i)$ in a sufficiently small neighborhood of  $x_0$, we obtain:
\[
\sum_{i=1}^n \{y_i - g_2^*(x_0)\}^2K_h(x_i - x)  = \sum_{i=1}^n \{y_i - g(x_0) - \lambda g(x_0) (x_i - x) -  \frac{1}{2}\lambda^2 g(x_0)(x_i - x)^2\}^2 K_h(x_i-x_0) \]
which leads to DE1-$2$ estimator
\[ \widehat{g}_2(x_0) = \frac{
\sum_{i=1}^n y_i(1+(x_i-x_0)\lambda + \frac{1}{2} (x_j -x)^2\lambda^2)K_h(x_i - x)}{\sum_{i=1}^n (1 + (x_i-x_0) + \frac{1}{2}(x_i-x_0)^2\lambda^2)^2 K_h(x_i-x_0)}.\] 
Since $g^{(p)}(x_0)=\lambda^p g(x_0)$, for $p=1,2,...,k$, the  $k^{th}$-degree estimator DE1-$k$ is obtained as
\begin{eqnarray}
\widehat{g}_k(x_0) 
&= \arg\min_{g(x_0)} \sum_{i=1}^n \left\{y_i- \sum_{p=0}^k \frac{1}{p!}(x_i-x_0)^p g^{(p)}(x_0)\right\}^2 K_h(x_i-x_0) \nonumber\\
&= \arg\min_{g(x_0)} \sum_{i=1}^n \left\{y_i-\sum_{p=0}^k \frac{1}{p!}(x_i-x_0)^p \lambda^{p}g(x_0)\right\}^2 K_h(x_i-x_0) \nonumber\\
&= \frac{\sum_{i=1}^n \{y_i \sum_{p=0}^k \frac{1}{p!}(x_i-x_0)^p \lambda^{p} \} K_h(x_i-x_0) }{\sum_{i=1}^n \{\sum_{p=0}^k \frac{1}{p!}(x_i-x_0)^p \lambda^{p} \}^2K_h(x_i-x_0)}.
%\label{equ:gkhat}
\end{eqnarray}
\section{Asymptotic Properties}

When performing the conditional asymptotic analysis of the $k^{th}$ degree estimators , we make the following assumptions for model (\ref{eqn:generallinear}):

(I) $g(x_0)$, the mean function, has a bounded and continuous $k+1^{th}$ derivative in a neighborhood of  $x_0$. 

(II) $f(x_0)$, the design density, is twice continuously differentiable and positive.

(III) $K_h(x)$, the kernel function, is a nonnegative, symmetric and bounded PDF with compact support, on the interval $[-h, h]$. The kernel function satisfies $\int_{-\infty}^{\infty}K_h(w)dw=1$, $R(K)=\int K_1^2(w)dw < \infty$, and has finite moments up to sixth order.   We will also use the notation $\mu_{k}=\int w^{k}K_1(w)dw$, and $R_k = \int w^{k} K_1^2(w)dw$.  

(IV) The error variance $\sigma^2(x)$ is a smooth function on $[a, b]$.  

\subsection{Asymptotic Conditional Bias and Variance Under Linear First Order DEs}

We will now summarize the results concerning the asymptotic conditional bias and variance of the DE-constrained estimators in the interior of the interval $[a,b]$ in the following two theorems. 

\textbf{Theorem 1 (Asymptotic Conditional Bias)} For regression model (\ref{eqn:generallinear}), under the assumptions (I) - (III), with $x_0 \in (a+h, b-h)$,  the DE1-$k$ estimator, $\hat{g}_k (x_0)$, has asymptotic conditional bias

\begin{equation}
\mathrm{Bias}(\widehat{g}_k (x_0)|x_1,...,x_n) = \frac{1}{(k+1)!}g^{(k+1)}(x_0)h^{k+1}\mu_{k+1}+o_p(h^{k+1}),  \quad k \quad \text{odd},
\end{equation}
%\begin{equation}
%\mathrm{Bias}(\widehat{g}_k (x_0)|x_1,...,x_n) = \frac{1}{(k+1)!}\lambda^{k+1}g(x_0)h^{k+1}\mu_{k+1}+o_p(h^{k+1}),  \quad k \quad \text{odd},
%\end{equation}
and 
\begin{equation}
\mathrm{Bias}(\widehat{g}_k (x_0)|x_1,...,x_n) =\left (\frac{g^{(k+2)}(x_0)}{(k+2)!}+\frac{g^{(k+1)}(x_0)}{(k+1)!}\frac{f'(x_0)}{f(x_0)}\right)h^{k+2}h^{k+2}\mu_{k+2}+o_p(h^{k+2}),  \quad k \quad \text{even},
\end{equation}

%\begin{equation}
%\mathrm{Bias}(\widehat{g}_k (x_0)|x_1,...,x_n) = \frac{1}{(k+1)!}\lambda^{k+1}g(x_0)h^{k+2}\mu_{k+2}(\frac{\lambda}{k+2}+\frac{f'(x_0)}{f(x_0)})+o_p(h^{k+2}),  \quad k \quad \text{even},
%\end{equation}
%The derivatives of $g(x_0)$, $g^{(k+1)}(x_0)=\sum_{l=0}^{k}\binom{k}{l}a^{l}(x_0)g^{(k-l)}(x_0)+b^{(k)}(x_0)$, and $g^{(k+2)}(x_0)=\sum_{l=0}^{k+1}\binom{k+1}{l}a^{l}(x_0)g^{(k+1-l)}(x_0)+b^{(k+1)}(x_0)$ .

\textbf{Theorem 2 (Asymptotic Conditional Variance)} Under the assumptions (I) through (IV), with $x_0 \in (a+h, b-h)$,  the DE1-$k$ estimator, $\hat{g}_k (x_0)$,  has asymptotic conditional variance
\begin{equation}
\mathrm{Var}(\widehat{g}_k (x_0)|x_1,...,x_n) =\frac{\sigma^2R(K)}{nhf(x_0)}+o_p\left(\frac{1}{nh}\right).
\end{equation}

%where $\frac{z-x}{h}=w$.

The proofs of Theorem 1 and 2  in the case of the local exponential model (\ref{equ:exp})  are provided in Appendix A. The proofs  for the model in the general linear form (\ref{eqn:generallinear}) are very similar to those for the local exponential model but with more
complicated expressions in the intermediate steps.  

\subsection{Asymptotically Optimal Bandwidths}

Bandwidth selection plays a crucial role in kernel-based regression models, directly influencing the quality and accuracy of the estimates. For differential equation-constrained regression models, bandwidth selection remains an essential consideration. Applying the conclusions from Theorems 1 and 2 in the case of the local exponential model (\ref{equ:exp}), we have the following corollary.

\textbf{Corollary 1 (Asymptotically Optimal Bandwidth)} Under the assumptions of model (\ref{equ:exp}), with $\lambda$ given, the asymptotically optimal bandwidths for  the DE1-$k$ are given by:
%\begin{equation}
%h_{o,k+2} =\Big(\frac{(k+3)(k+1)}{\lambda^4}h_{o,k}^{2k+3}\Big)^{1/(2k+7)}, \quad k \quad \text{odd},
%\label{equ: oddh}
%\end{equation}
\begin{equation}
h_{o,k}^{2k+3}= \frac{\sigma^2R(K)\{(k+1)!\}^2}{nf(x_0)e^{2\lambda x}\lambda^{2k+2}(2k+2)\mu_{k+1}^2}
\mbox{ and }
h_{o,k+2} =\Big(\frac{(k+3)(k+1)}{\lambda^4}h_{o,k}^{2k+3}\Big)^{1/(2k+7)},  \ \ \ \ k \mbox{ odd}
\label{equ: oddh}
\end{equation}
and
\begin{equation}
h_{o,k}^{2k+5}= \frac{\sigma^2R(K)\{(k+1)!\}^2}{nf(x_0)e^{2\lambda x}\lambda^{2k+2}(2k+4)\mu_{k+2}^2\left(\lambda+\frac{f'(x_0)}{f(x_0)}\right)^2}
\mbox{ and }
h_{o,k+2} =\Big(\frac{(k+2)^3}{(k+4)\lambda^4}h_{o,k}^{2k+5}\Big)^{1/(2k+9)},  \ \ \ \ k \mbox{ even}. 
\label{equ: evenh}
\end{equation}

The proof is provided in Appendix A.

\textbf{Remark} (Application of Corollary 1): 

To find the optimal bandwidth for even $k$, denoted as $h_{o,k}$, we can begin by selecting an initial bandwidth $h_{o,0}$ for the local constant estimator. This can be determined using (\ref{equ: evenh}) with $k = 0$.
\begin{equation}
h_{o,0}^5= \frac{\sigma^2R(K)}{4nf(x_0)e^{2\lambda x}\lambda^2 \left(\lambda+\frac{f'(x_0)}{f(x_0)}\right)^2} \nonumber
\end{equation}
By substituting $k = 2$ in   (\ref{equ: evenh}), we obtain the optimal bandwidth for the second-degree DE-constrained regression approach. 
\begin{equation}
h_{o,2} = \left(\frac{2}{\lambda^4}h_{o,0}^5\right)^{1/9}. \nonumber 
\end{equation}
Successive applications of the same formula will yield the optimal bandwidths for $k =  4, 6, 8 \dots$ (even degrees). 
The corresponding bandwidths for odd degrees of $k$ can be determined using formula (\ref{equ: oddh}).
Example 3 of the next section will illustrate an application for Corollary 1 when the bandwidths are chosen according to the asymptotically optimal criterion..

\subsection{Theoretical Comparison of Estimators}
In this section, we will compare the asymptotic properties for different estimators, assuming the correct model.
We will compare our DE-constrained estimators with the conventional local polynomial regression estimators and \citet{he2009double}'s double-smoothing estimator. 

Under the DE-constrained local exponential growth model (\ref{equ:exp}),  $g^{(k)}(x_0)=\lambda^k g(x_0)$. Table \ref{table:asymptotic} provides  a summary comparison of asymptotic properties of various estimators. The order follows the pattern from the largest to smallest bias magnitude, approximately. 

The double-smoothed estimator (DS) was proposed by \citet{he2009double}.   Related notation is as follows. 
$B(x_0)=(\mu_2^2-\mu_4)/4[g''(x_0)f''(x_0)/f(x_0)+2(g^{(3)}(x_0)f'(x_0)/f(x_0))+g^{(4)}(x_0)] = (\mu_2^2-\mu_4)/4[\lambda^2g(x_0)f''(x_0)/f(x_0)+2(\lambda^3g(x_0)f'(x_0)/f(x_0))+\lambda^4g(x_0)] $.  $V=\int \{(K*K)(v)-(L*L)(v)/\mu_2\}^2dv$, where $L(u)=uK_1(u)$, and  $v_k = \int w^kK_1^2(w)dw$, for $k=0,1,2,...$.

\begin{table}[h!]
  \begin{center}  
    \begin{tabular}{c|c|c} % <-- Alignments: 1st column left, 2nd middle and 3rd right, with vertical lines in between
      \hline
      \textbf{Method} & \textbf{Asymptotic bias in interior} & \textbf{Asymptotic variance}\\
      \hline
      NW & $\frac{1}{2}\left\{\lambda^2g(x_0)+2\lambda\frac{g(x_0)f'(x_0)}{f(x_0)}\right\}h^2 \mu_2+o_p(h^2)$  & $\frac{1}{nh}\frac{\sigma^2}{f(x_0)}v_0+o_p(\frac{1}{nh})$ \\
      LL & $\frac{1}{2}\lambda^2g(x_0)h^2 \mu_2 +o_p(h^2)$ & $\frac{1}{nh}\frac{\sigma^2(x_0)}{f(x_0)}v_0+o_p(\frac{1}{nh})$\\
      DE1-1 & $\frac{1}{2}\lambda^2g(x_0)h^2\mu_2+o_p(h^2)$ & $\frac{1}{nh}\frac{\sigma^2}{f(x_0)}v_0+o_p(\frac{1}{nh})$ \\
      DE1-2  & $\frac{1}{6}\lambda^3g(x_0)h^4\mu_4\left(\lambda + \frac{f'(x_0)}{f(x_0)}\right)\mu_4+ o_p(h^4)$ & $\frac{1}{nh}\frac{\sigma^2}{f(x_0)}v_0+o_p(\frac{1}{nh})$\\  
      LQ  & $\frac{1}{24}\frac{\mu_2\mu_6-\mu_4^2}{\mu_2^2-\mu_4}\left\{\lambda^4g(x_0)+4\lambda^3\frac{g(x_0)f'(x_0)}{f(x_0)}\right\}h^4+o_p(h^4)$  & $\frac{1}{nh}\frac{\sigma^2}{f(x_0)}\frac{\mu_4^2v_0-2\mu_2\mu_4v_2+\mu_2^2v_4}{(\mu_2^2-\mu_4)^2}+o_p(\frac{1}{nh})$\\ 
      LC & $ \frac{1}{24}\frac{\mu_2\mu_6-\mu_4^2}{\mu_2^2-\mu_4}\lambda^4g(x_0)h^4+o_p(h^4)$  & $\frac{1}{nh}\frac{\sigma^2}{f(x_0)}\frac{\mu_4^2v_0-2\mu_2\mu_4v_2+\mu_2^2v_4}{(\mu_2^2-\mu_4)^2}+o_p(\frac{1}{nh})$\\ 
      DS & $h^4B(x_0)+o_p(h^4)$ & $\frac{1}{nh}\frac{\sigma^2}{f(x_0)}V+o_p(\frac{1}{nh})$\\       
      DE1-3 &$\frac{1}{24}\lambda^4g(x_0)h^4\mu_4+o_p(h^4)$ & $\frac{1}{nh}\frac{\sigma^2}{f(x_0)}v_0+o_p(\frac{1}{nh})$\\
     \hline      
    \end{tabular}    
    \caption{Summary of asymptotic conditional bias and variance for the NW (local constant), LL (local linear), LC (local cubic), DS (double smoothing), DE1-1, DE1-2, and DE1-3 estimators.  Here, $g(x_0) = g(0)\mbox{e}^{\lambda x_0}$.}
\label{table:asymptotic} 
  \end{center}   
\end{table}

We observe that the higher-degree estimator has a lower bias than the conventional local polynomial estimators with the same degree, while the variance stays the same.  Therefore,
if the DE model is correct, we would prefer a higher-degree DE-constrained estimator.  

The ratio of the variance expressions is independent of $\sigma^2$.  Further direct analysis of the ratio is difficult, but simulation provides some insight.  
In a small study, we assume $\lambda = 1$ and take a sample of size 10 with a random uniform sampling design.  We then calculated the ratio of the variances at each design point using a Gaussian kernel. 
We found that the average of the ratio values is $0.998$ with a range of values is (0.893, 1.096).  These results are typical.  Thus, we conclude that the variance behavior of the first-order approximation to the exponential growth model is quite similar to that of local constant regression.
    
\section{Simulation Study}

We now study the method numerically by simulating 1000 sets of data with sample sizes 30 and 100, assuming pure exponential growth: 
$$g(x) = g(0) \mbox{e}^{\lambda x}, $$
where we take the initial condition to be $g(0) = 1$.   
In this simulation, we assume that the noise follows a normal distribution with zero mean and a standard deviation of $\sigma$.

We present the simulation results for $\lambda$ = 1 and $\sigma$ = 0.1, 0.2, and 0.3. The study design includes both a uniform design,  $x \sim$ U[0,1], and a sparse design, $x \sim$ Beta(1, 0.5). These two designs aim to evaluate the performance of the DE-constrained estimators under varying conditions.  

Additional simulation studies, detailed in the appendices, explore other sparse designs, such as Beta(1.25, 1), Beta(2, 5), Beta(5, 1) and Beta(3, 3). For these cases, $\lambda$ values of 0.5, 1, and 2 are considered. We also examine models with non-normally distributed noise, such as t(4) and Laplace(0, 1). The results of simulation studies are presented in Appendix B.

For each simulated dataset, we apply the following estimators: NW (local constant), LL (local linear), LQ (local quadratic), LC (local cubic), DE1-1 (first-degree DE-constrained exponential growth approximation), DE1-2 (second-degree DE-constrained exponential growth approximation), DE1-3 (third-degree DE-constrained exponential growth approximation), DE1-4 (fourth-degree DE-constrained exponential growth approximation), DE1-5 (fifth-degree DE-constrained exponential growth approximation), and NLS (the exponential growth model fit by nonlinear least-squares).
The nonlinear least-squares result serves as a gold standard here.  In practice, the solution of the DE may not be available in closed-form so the nonlinear regression estimator will not always be an option.  

\begin{figure}
        \centering
        \vspace{-20mm}
        \includegraphics[width=1\textwidth]{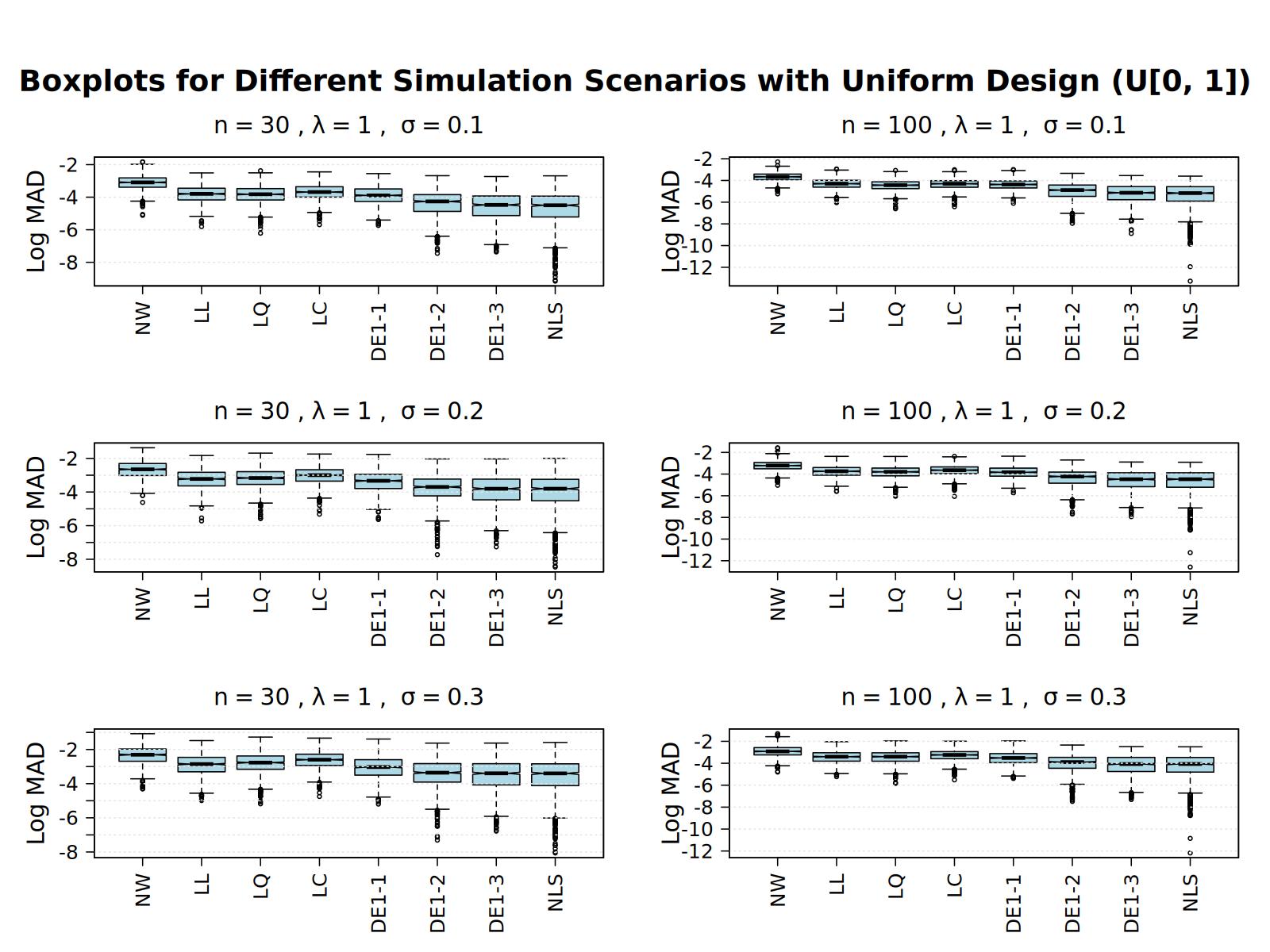}
        \vspace{-9mm}
        \caption{MAD (Median Absolute Deviation) distributions (on the log scale) for the estimation methods
applied to simulated data with a uniform design x $\sim$ U[0, 1]. Sample size n = 30, 100. $\lambda$ = 1. Note: $\mathrm{MAD}=\mathrm{median}(|y_i-\hat{y}_i|).$}
        \label{fig:unif}
        \end{figure}

\begin{figure}
        \centering
        \vspace{-20mm}
        \includegraphics[width=1\textwidth]{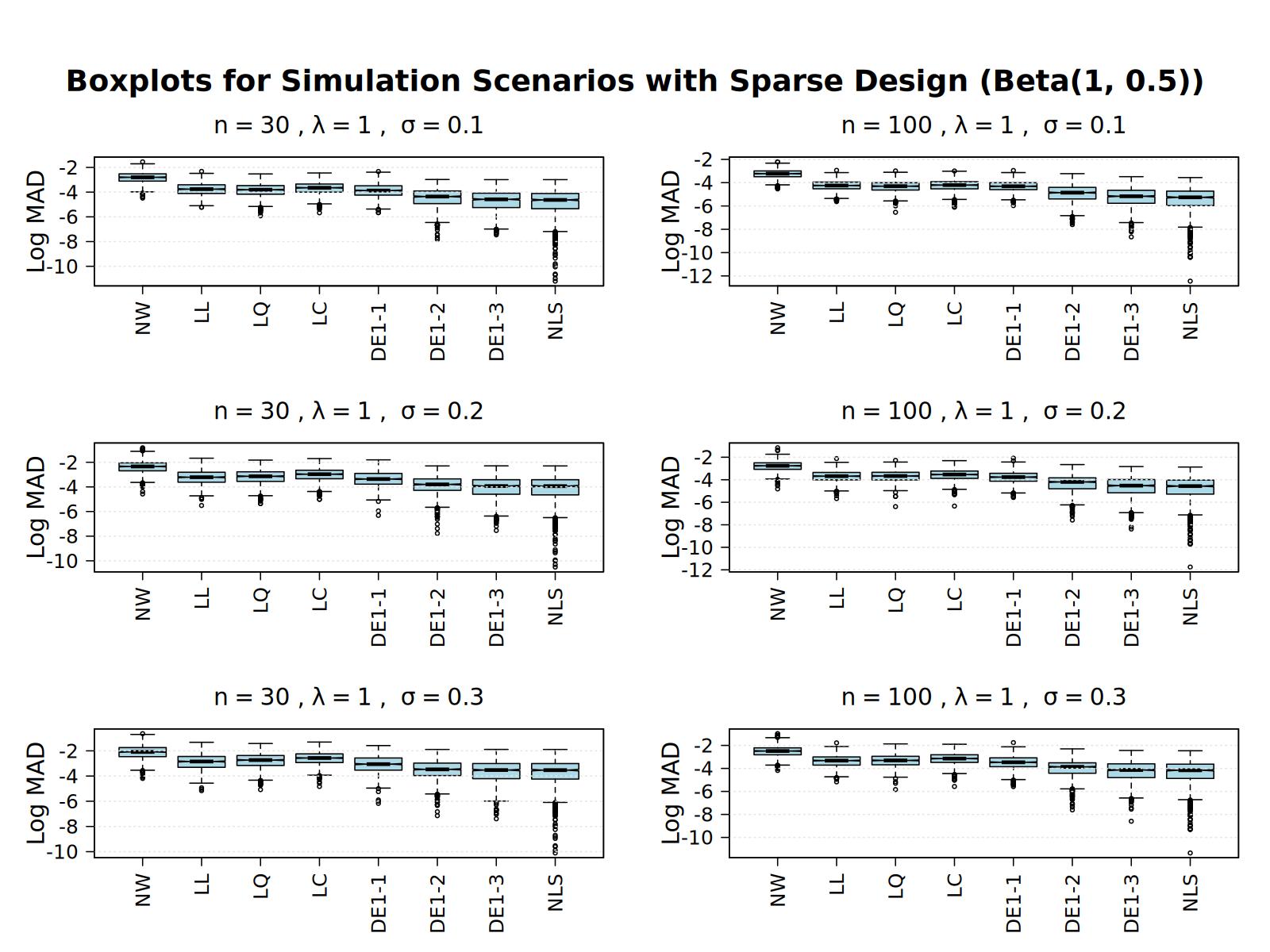}
        \vspace{-9mm}
        \caption{MAD (Median Absolute Deviation) distributions (on the log scale) for the estimation methods
applied to simulated data with a sparse design x $\sim$ Beta(1, 0.5). Sample size n = 30, 100. $\lambda$ = 1. Note: $\mathrm{MAD}=\mathrm{median}(|y_i-\hat{y}_i|).$}
        \label{fig:Spa}
        \end{figure}

 Bandwidths are selected by leave-one-out cross-validation. The simulation results, DE1-3, DE1-4, and DE1-4 are very similar. Therefore, we exclude the results of DE1-4 and DE1-5 in Figure \ref{fig:unif}, Figure \ref{fig:Spa}, Table \ref{table:Uni}, and Table \ref{table:Spa}.

 In the simulation study for Examples 1 and 2 (below), we employed the cross-validation method to select bandwidths for each DE-constrained estimator, which works well for small sample sizes.

For larger datasets, such as those with 10,000 observations, bandwidths can be effectively selected using Corollary 1. Example 3 includes a simulation study on optimal bandwidth selection.

Our basis for comparing the methods is the \textit{median absolute deviation} (MAD), which is calculated by taking the absolute value of the difference between the fitted values and the corresponding true model values, and finding the median. That is,
\[
\text{MAD} = \text{median}(|y_i - \hat{y}_i|),
\]
where \( y_i \) is the true model value, and \( \hat{y}_i \) is the fitted value for the corresponding data point. We then report the average of the MADs for the set of 1000 simulated datasets. Additionally, we display the distribution of MAD values in boxplots for each method.

%We also explored the issue of model misspecification. However, it is not the main aim of this paper.  This will be touched on in the last section.

\noindent
\textbf{Example 1. Uniform Sampling Design}

In our first set of simulations, we consider a random uniformly distributed design for the covariate on the interval $[0, 1]$.  
Boxplots are shown in Figure \ref{fig:unif}, illustrating the distributions (on the log scale) of Median Absolute Deviation (MAD) for the 8 different estimation methods described earlier, applied to simulated data with a uniform design. Table \ref{table:Uni} presents the means and standard errors of MAD for each method and for the two sample sizes, n = 30 and 100. 

\begin{table}
    \centering
   \vspace{-3mm}
    \scalebox{0.9}{
    \begin{tabular}{lccc}
           \toprule
        & \multicolumn{3}{c}{Sample size n = 30, Mean (Standard Error)} \\
\hline
        & $\sigma$ = 0.1& $\sigma$ = 0.2 & $\sigma$ = 0.3 \\
\hline
        \midrule
NW      & 48.96 (0.69) & 78.91 (1.29) & 109.31 (1.78) \\  %New
LL        & 24.86 (0.39) & 44.88 (0.75) & 64.08 (1.10)\\ %New
LQ       & 24.10 (0.38) & 47.12 (0.79) & 70.32 (1.19) \\ %New
LC       & 27.15 (0.39) & 53.61 (0.78) & 80.03 (1.19) \\ %New
DE1-1 & 23.39 (0.39) & 40.97 (0.76) & 56.78 (1.12)\\ %New
DE1-2 & 15.77 (0.32) & 29.30 (0.63) & 42.63 (0.94)\\ %New
DE1-3 & 13.70 (0.32) & 27.26 (0.64) & 40.87 (0.95) \\ %new
NLS    & 13.39 (0.32) & 26.78 (0.64) & 40.18 (0.95) \\ %new
\\
        \midrule
       & \multicolumn{3}{c}{Sample size $n = 100$, Mean (Standard Error)} \\
\hline
        & $\sigma = 0.1$ & $\sigma = 0.2$ & $\sigma = 0.3$ \\
\hline
        \midrule
       NW  &  27.11 (0.34)  &  44.17 (0.71)  &  61.92 (1.09)  \\ 
   LL  &  14.72 (0.21)  &  26.24 (0.41)  &  36.91 (0.61)   \\ 
   LQ  &  13.01 (0.20)  &  24.88 (0.41)  &  36.78 (0.62)  \\ 
   LC  &  14.41 (0.21)  &  28.13 (0.42)  &  41.71 (0.63)   \\ 
   DE1-1  &  13.93 (0.21)  &  24.41 (0.41)  &  33.92 (0.61)   \\ 
   DE1-2  &  8.70 (0.18)  &  16.24 (0.34)  &  23.33 (0.51)   \\ 
   DE1-3  &  7.31 (0.17)  &  14.28 (0.34)  &  21.34 (0.52)  \\ 
  NLS  &  7.04 (0.17)  &  14.08 (0.34)  &  21.12 (0.52)  \\ 
        \bottomrule
    \end{tabular}
    }
    \caption{Means and Standard Errors for Median Absolute Deviation (MAD) multiplied by 1000. The top section shows the mean values (Standard Error values) for sample size n = 30, and the bottom section shows the values for sample size n = 100. Both results are based on 1000 simulation runs under the uniform sampling design U[0, 1] with $\lambda$ = 1, noise $\varepsilon \sim$ N(0, $\sigma$), where $\sigma$ = 0.1, 0.2, 0.3.}
    \label{table:Uni}
 \vspace{-14mm}
\end{table}

As expected, 
we see that the solution of the differential equation is the most accurate estimator, i.e., the NLS estimator, followed by the fifth degree DE approximation.  The accuracy of the methods decreases as the degree decreases.  The completely nonparametric methods are not competitive with the DE-constrained methods in this case.

According to the boxplots and tables for the local growth model, we find that the highest-degree estimator (DE1-5) has the lowest bias.  For  model selection, a higher-degree DE-constrained regression is preferable.

\noindent
\textbf{Example 2. Sparse Sampling Design}

In our second set of simulations, we consider a random beta distributed design for the covariate on the interval $[0, 1]$.  Shape parameters for the beta distribution were taken to be 1.0 and 0.5.  This induces sparsity near the origin.  

Boxplots of the MAD distributions for the various methods applied to the sparse designs are displayed in Figure \ref{fig:Spa}.  Table \ref{table:Spa} provides analogous information to that of Tables \ref{table:Uni}.  

According to the boxplots and tables for the local growth model, we find that the completely nonparametric methods are not competitive with the DE-constrained methods.  Furthermore, under sparse design, the highest-degree estimator (DE1-3) still has the lowest bias.  For model selection considerations, a higher-degree DE-constrained estimator is a better choice. 

\begin{table}[ht]
    \centering
 \vspace{-3mm}
    \scalebox{0.9}{
    \begin{tabular}{lccc}
        \toprule
        & \multicolumn{3}{c}{Sample size n = 30, Mean (Standard Error)} \\
\hline
        & $\sigma$ = 0.1& $\sigma$ = 0.2 & $\sigma$ = 0.3   \\
\hline
        \midrule
   NW    & 64.58 (0.86) & 105.20 (1.73) & 140.47 (2.45)  \\ %New
  LL       & 26.23 (0.43) & 46.51 (0.84)  & 66.00 (1.22) \\ %New
  LQ      & 24.48 (0.40) & 47.78 (0.82) & 71.32 (1.24) \\ %New
  LC       & 28.14 (0.42) & 55.36 (0.85) & 82.50 (1.29) \\  %New
  DE1-1 & 23.96 (0.41) & 41.28 (0.79) & 56.21 (1.12)\\ %new
  DE1-2 & 14.43 (0.29) & 26.01 (0.54) & 37.36 (0.79)\\ %new
  DE1-3 & 11.87 (0.27) & 23.36 (0.53) & 34.89 (0.79)\\ %new
  NLS    & 11.39 (0.27) & 22.77 (0.53) & 34.16  (0.80)\\ %new

\\
        \midrule
        & \multicolumn{3}{c}{Sample size n = 100, Mean (Standard Error)} \\
\hline
        & $\sigma$ = 0.1 & $\sigma$ = 0.2 & $\sigma$ = 0.3 \\
\hline
        \midrule
  NW     & 41.40 (0.49) & 66.80 (0.94) & 90.29 (1.45)\\ %New
  LL        & 15.54 (0.21) & 27.96 (0.43) & 39.64 (0.65)\\ %New
  LQ       & 14.54 (0.21) & 27.80 (0.44) & 40.80 (0.67)\\ %New
  LC        & 15.90 (0.22) & 31.21 (0.45) & 46.58 (0.68) \\  %New
  DE1-1  & 14.77 (0.21) & 25.80 (0.44) & 35.71 (0.66)\\ %new
  DE1-2  & 8.85 (0.18) & 16.05 (0.32) & 22.58 (0.46)\\ %new
  DE1-3  & 6.84 (0.15) & 13.12 (0.30) & 19.37 (0.45)\\ %new
  NLS     & 6.23 (0.15) & 12.46 (0.30) & 18.69 (0.46)\\ %new
        \bottomrule
    \end{tabular}
    }
    \caption{Means and Standard Errors for Median Absolute Deviation (MAD) multiplied by 1000. The top section shows the mean values (Standard Error values) for sample size n = 30, and the bottom section shows the values for sample size n = 100. Both results are based on 1000 simulation runs under the sparse sampling design Beta(1, 0.5) with $\lambda$ = 1, noise $\varepsilon \sim$ N(0, $\sigma$), where $\sigma$ = 0.1, 0.2, 0.3.}
 \vspace{-4mm}
    \label{table:Spa}
\end{table}

\begin{figure}
        \centering
        \vspace{-4mm}
        \includegraphics[width=1\textwidth]{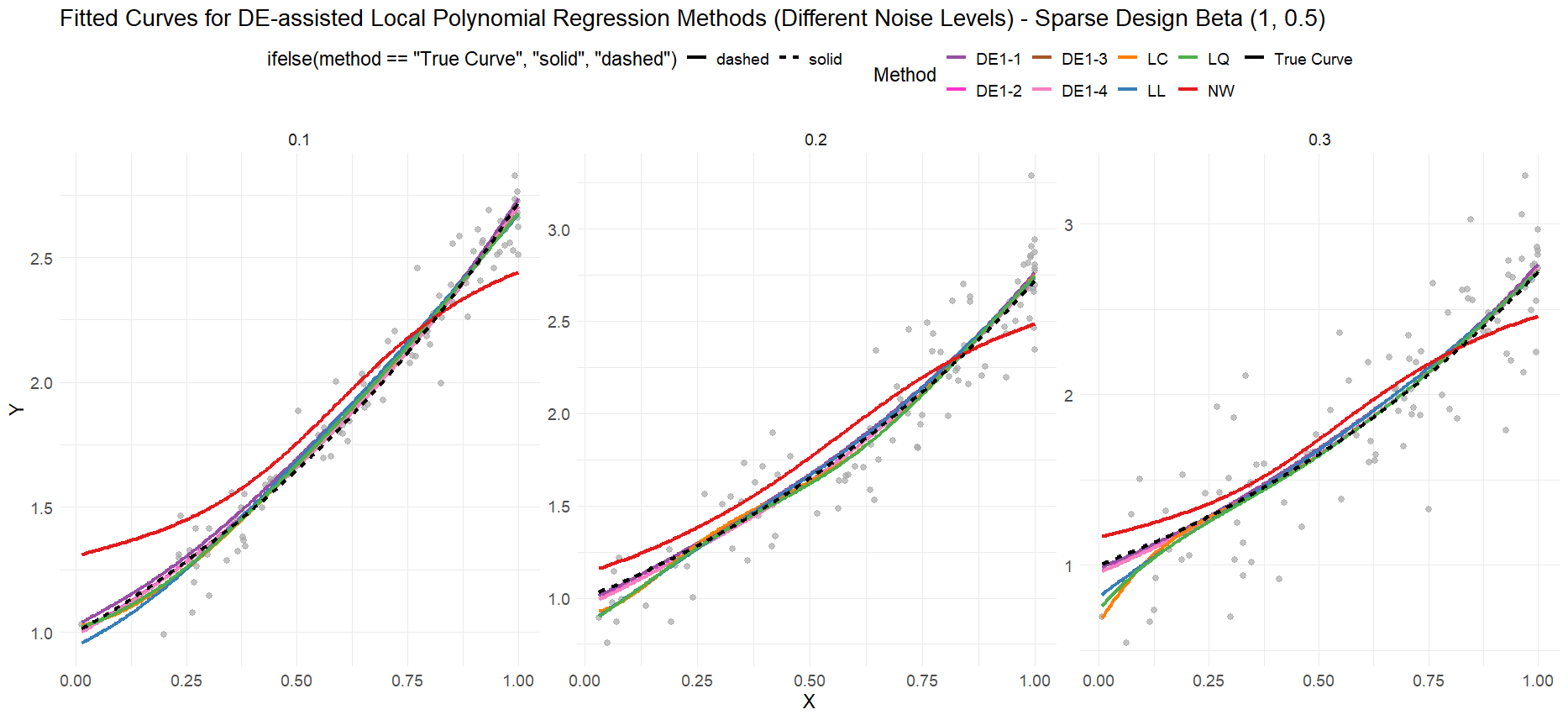}
        \vspace{-4mm}
        \caption{Fitted curves for DE-constrained regression and local polynomial regression methods applied to simulated data with a sparse design Beta(1, 0.5) at different noise levels, $\sigma$ = 0.1, 0.2, 0.3. Sample size n = 100. $\lambda$ = 1. }
        \label{fig:SpaFitting}
        \end{figure}

\begin{figure}
        \centering
        \vspace{-4mm}
        \includegraphics[width=1\textwidth]{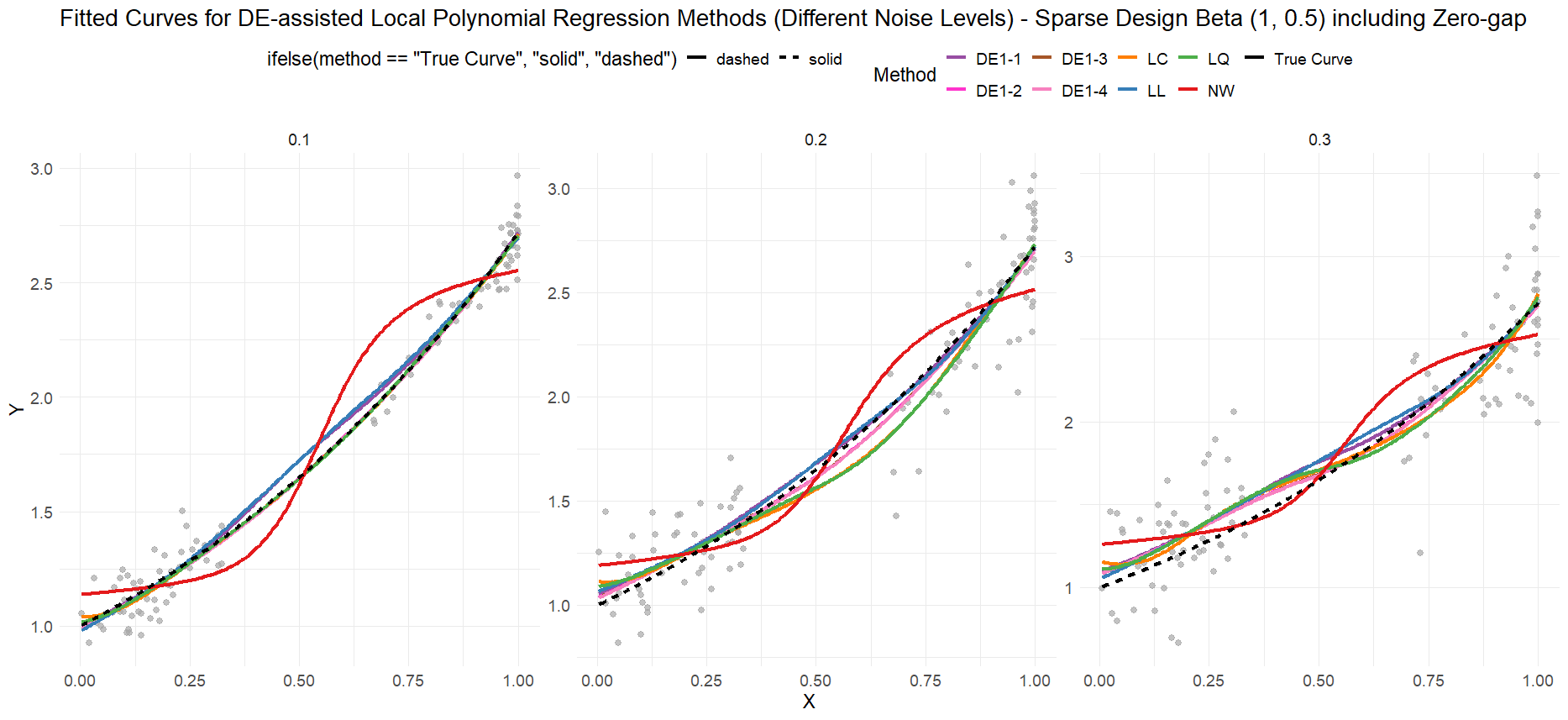}
        \vspace{-4mm}
        \caption{Fitted curves for DE-constrained regression and local polynomial regression methods applied to simulated data with a sparse design Beta(1, 0.5) including Zero-gap at different noise levels, $\sigma$ = 0.1, 0.2, 0.3. Sample size n = 100. $\lambda$ = 1. }
        \label{fig:SpaFittingMix}
        \end{figure}

To further validate the good performance of DE-constrained estimates, and to get a visual impression of the magnitude of the effect achieved, we conducted some additional one-off simulations and plotted the results in Figures \ref{fig:SpaFitting} and \ref{fig:SpaFittingMix}. 
Figure \ref{fig:SpaFitting} shows a plot of fitted curves for DE-constrained regression and conventional local polynomial regression methods under a Beta(1, 0.5) design with different noise levels. All the fitted curves of DE-constrained estimators are closer to the true curve than local polynomial estimators. 
We also designed a new type of sparse data with Beta(1, 0.5) and a zero-gap within the data points. The result in Figure \ref{fig:SpaFittingMix} demonstrates the advantages of DE-constrained methods over local polynomial methods.

\noindent
\textbf{Example 3. Large Sample Comparisons}

By comparing the log-transformed Mean Squared Error (MSE) as a function of the covariate  $x_0$, this example evaluates and visualizes the performance of different regression models, DE1-0 (i.e. Local Constant), DE1-$k$, ($k$ = 1, ..., 7), and LL (Local Linear) for an exponential growth function with varying noise levels ($\sigma$ = 0.1, 0.2, 0.3) and growth rates ($\lambda$ = 0.5, 1, 2). 

% The NLS (nonlinear least-squares) regression estimator is included as a ``gold standard''.  It is not possible for any nonparametric
%or semiparametric method to achieve better performance than this.  

The sample size considered here is $n = 10000$, and the sparse design is based on quantiles of a Beta distribution with shape parameters 1 and 0.5.  Two additional examples are provided in Appendix B.  

 The results in Figure \ref{fig:logMSEBeta105} indicate that the higher-degree DE-constrained methods exhibit lower Log(MSE) values, with all DE1-$k$, $k\ge2$, outperforming the local constant and local linear methods. However, DE1-1 performs worse than local linear methods at the boundaries.   Boundary bias is negligible for
 DE methods of degree greater than 2.  Furthermore, when $\lambda$ is small, the 
 higher degree DE methods perform almost as well as the NLS method.  For larger values of $\lambda$, there is a gap between NLS and the DE methods for  small values of  $x_0$, but NLS performance degrades quickly as  $x_0$ increases, while the DE methods continue to perform consistently across the entire range of  $x_0$ values.   Similar remarks apply to the designs considered in Appendix B.  

         \begin{figure}
        \centering
        \vspace{-4mm}
        \includegraphics[width=1\textwidth]{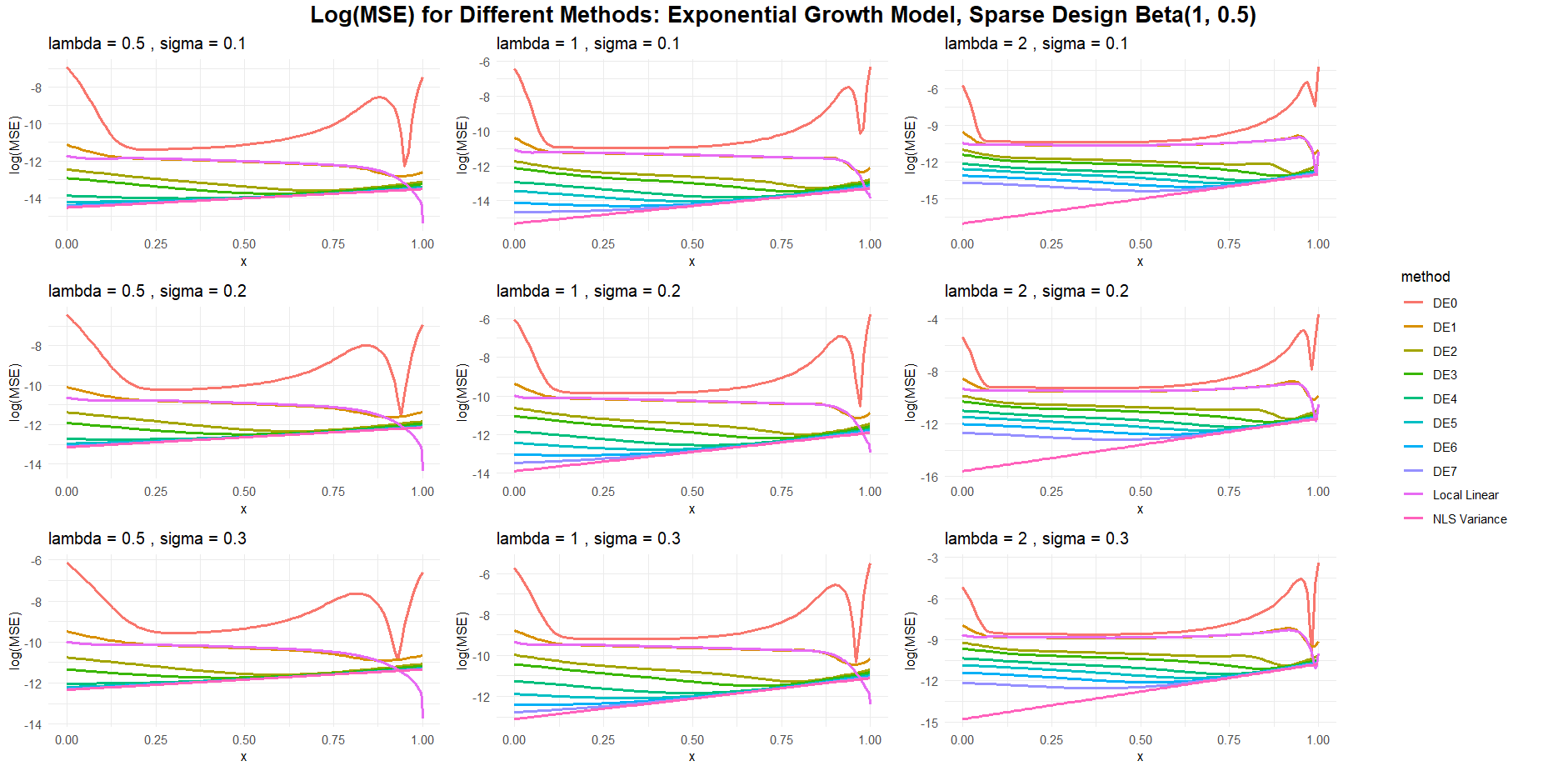}
        \vspace{-4mm}
        \caption{Log(MSE) for DE-constrained regression and local polynomial regression methods applied to simulated data with a sparse design Beta(1, 0.5) at different noise levels, $\sigma$ = 0.1, 0.2, 0.3. Sample size n = 10000. $\lambda$ = 0.5, 1, 2. }
        \label{fig:logMSEBeta105}
        \end{figure}

 \section{Discussion}

Differential equation information can enhance local polynomial regression estimates. The  method 
studied in this paper 
is simple, requiring minimal parameter tuning. It doesn't require solving the differential equation, and it is only slightly more complicated to implement than local constant regression.  As a result, it is both simple and computationally efficient.

The number of derivatives needed depends on how reasonable the differential equation (DE) is for the given problem. If the DE seems less reliable, fewer terms  and a smaller bandwidth $h$ should be used. If the DE is nonlinear, weighted nonlinear least-squares may be needed for estimation. Global parameters, such as $\lambda$ in the growth model, also need to be estimated. 

The method reduces bias compared to local polynomial regression with same degree, both in the interior and at the data boundaries. It is relatively robust to model misspecification and performs particularly well with sparse designs compared to other methods.  

Bias reduction can still be achieved even if the parameters are estimated.   Bias order remains $O(h^2)$ as long as parameter estimates
converge at rate $n^{1/2}$ (the typical parametric convergence rate) and even if parameter estimates converge at rate $n^{2/5}$ (the typical nonparametric convergence rate).  

We have focussed on a simple case, the exponential model, as a proof of concept. In the future, study of more complicated models will be of interest. Model misspecification is an area we plan to investigate further in another paper, focusing on misspecification analysis for DE-constrained estimators.

The DE-constrained regression approach shows good performance with real datasets. For this paper, we used the mouse tumor data to illustrate the estimation method.   An important aspect of this class of estimators is their robustness to model misspecification.  We have conducted some simulation analysis and have found that low order DE-constrained methods enjoy a certain degree of robustness.  The details of this work will be published in a separate paper in the near future. 

%In future studies, we will use the "attenu" dataset (the Joyner-Boore Attenuation Data) to explore model misspecification further.
This paper mainly focused on the first-order DE-constrained regression model. Further exploration of higher-order models is planned, as well as nonlinear differential equation models.  

%and the classic "faithful" dataset (Old Faithful Geyser Data) will be used to study the second-order DE-constrained regression model.

{\bf Acknowledgements} 

This research has been supported in part by a grant from the Natural Sciences and Engineering Research Council of Canada (NSERC).   We are grateful to two anonymous reviewers for helpful comments and suggestions.  

We also acknowledge the assistance of ChatGPT for grammar and punctuation checks in preparing this manuscript.

\bibliographystyle{apalike}
\bibliography{references}

% Appendices
\begin{appendices}

\section{Proof of Theorems}

\subsection{Proof of Theorem 1}
The asymptotic conditional bias of the $k^{th}$-degree estimator $\hat{g}_k (x_0)$ is given by

\begin{equation}
\begin{split}
\mathrm{Bias}(\widehat{g}_k (x_0)|x_1,...,x_n) &=\mathrm{E}[\hat{g}_k(x_0)|x_1,...x_n]-g(x_0) \\
& \approx \frac{1}{(k+1)!}\lambda^{k+1} g(x_0) \frac{\sum_{i=1}^n (x_i-x_0)^{k+1}\{\sum_{p=0}^k \frac{1}{p!}(x_i-x_0)^p \lambda^{p}\}K_h(x_i-x_0) }{\sum_{i=1}^n\{\sum_{p=0}^k \frac{1}{p!}(x_i-x_0)^p \lambda^{p} \}^2K_h(x_i-x_0)}\\
& \approx \frac{1}{(k+1)!}\lambda^{k+1} g(x_0)  \frac{\int_a^b (z-x)^{k+1}\{\sum_{p=0}^k \frac{1}{p!}(z-x)^p \lambda^{p}\}f(z)K_h(z-x)dz}{\int_a^b\{\sum_{p=0}^k \frac{1}{p!}(z-x)^p \lambda^{p} \}^2f(z)K_h(z-x)dz}
\end{split}
\end{equation}
 Let $\frac{z-x}{h}=w$. Then the integral in the numerator becomes
\begin{equation*}
\begin{split}
& \int_a^b (z-x)^{k+1}\left\{\sum_{p=0}^k \frac{1}{p!}(z-x)^p \lambda^{p}\right\}f(z)K_h(z-x)dz \\
&= \int (hw)^{k+1}\left\{\sum_{p=0}^k \frac{1}{p!}(hw)^p \lambda^{p}\right\}f(x+hw)K(w)dw \\
&= \int (hw)^{k+1}\left\{\sum_{p=0}^k \frac{1}{p!}(hw)^p \lambda^{p}\right\}(f(x_0)+hwf'(x_0))K(w)dw \\
&= \int\left \{ \sum_{p=0}^k\frac{1}{p!}h^{p+k+1}w^{p+k+1} \lambda^{p}f(x_0)+\sum_{p=0}^k\frac{1}{p!}h^{p+k+2}w^{p+k+2} \lambda^{p}f'(x_0) \right\}K_1(w)dw. 
\end{split}
\end{equation*}
The denominator can be approximated by
\begin{equation*}
\begin{split}
& \int_a^b\left\{\sum_{p=0}^k \frac{1}{p!}(z-x)^p \lambda^{p}\right \}^2f(z)K_h(z-x)dz \\
& = \int \left\{\sum_{p=0}^k \frac{1}{p!}(hw)^p \lambda^{p} \right\}^2f(x+hw)K_1(w)dw \\
& \approx f(x_0)
\end{split}
\end{equation*}
when $x_0 \in (a+h, b-h)$, and when $k$ is odd, the asymptotic conditional bias of $\widehat{g}_k (x_0)$ is
\begin{equation}
\mathrm{Bias}(\widehat{g}_k (x_0)|x_1,...,x_n) = \frac{1}{(k+1)!}\lambda^{k+1}g(x_0)h^{k+1}\mu_{k+1}+o_p(h^{k+1})
\end{equation}
and when $k$ is even, 
\begin{equation}
\mathrm{Bias}(\widehat{g}_k (x_0)|x_1,...,x_n) = \frac{1}{(k+1)!}\lambda^{k+1}g(x_0)h^{k+2}\mu_{k+2}(\frac{\lambda}{k+2}+\frac{f'(x_0)}{f(x_0)})+o_p(h^{k+2}). 
\end{equation}
\subsection{Proof of Theorem 2}

Assume $x_0 \in (a+h, b-h)$.  
The asymptotic conditional variance of the $k^{th}$-degree estimator (DE1-$k$ estimator) is then
\begin{align}
\mathrm{Var}(\widehat{g}_k (x_0)|x_1,...,x_n)
&= \frac{\sum_{i=1}^n  \mathrm{Var}(y_i|x_1,...,x_n) \{\sum_{p=0}^k \frac{1}{p!}(x_i-x_0)^p \lambda^{p} \}^2 K_h^2(x_i-x_0) }{ \{\sum_{i=1}^n \{\sum_{p=0}^k \frac{1}{p!}(x_i-x_0)^p \lambda^{p} \}^2K_h(x_i-x_0) \}^2} \nonumber \\
&= \frac{\sum_{i=1}^n  \sigma^2 \{\sum_{p=0}^k \frac{1}{p!}(x_i-x_0)^p \lambda^{p} \}^2 K_h^2(x_i-x_0) }{ \{\sum_{i=1}^n \{\sum_{p=0}^k \frac{1}{p!}(x_i-x_0)^p \lambda^{p} \}^2K_h(x_i-x_0) \}^2} \nonumber \\
& \approx \frac{1}{n} \frac{\int_a^b  \sigma^2 \{1+\sum_{p=1}^k \frac{1}{p!}(z-x)^p \lambda^{p} \}^2 K_h^2(z-x)f(z)dz }{ \{\int_a^b \{1+\sum_{p=1}^k \frac{1}{p!}(z-x)^p \lambda^{p} \}^2K_h(z-x) f(z)dz\}^2} \nonumber \\
&=\frac{1}{nh} \frac{\int  \sigma^2 \{1+\sum_{p=1}^k \frac{1}{p!}(hw)^p \lambda^{p} \}^2 K^2(w)f(x+hw)dw }{ \{\int \{1+\sum_{p=1}^k \frac{1}{p!}(hw)^p \lambda^{p} \}^2K(w) f(x+hw)dw\}^2} \nonumber \\
&\approx \frac{\sigma^2R(K)}{nhf(x_0)}+o_p\left(\frac{1}{nh}\right)
\end{align}
where $\frac{z-x}{h}=w$.
If $\sigma^2$ varies with  $x$, that is, the conditional variance $\mathrm{Var}(y_i|x_1,...,x_n) = \sigma^2(x_i)$, then the asymptotic conditional variance of $\widehat{g}_k (x_0)|x_1,...,x_n)$
is given by
\begin{align}
\mathrm{Var}(\widehat{g}_k (x_0)|x_1,...,x_n)
&= \frac{\sum_{i=1}^n  \mathrm{Var}(y_i|x_1,...,x_n) \{\sum_{p=0}^k \frac{1}{p!}(x_i-x_0)^p \lambda^{p} \}^2 K_h^2(x_i-x_0) }{ \{\sum_{i=1}^n \{\sum_{p=0}^k \frac{1}{p!}(x_i-x_0)^p \lambda^{p} \}^2K_h(x_i-x_0) \}^2} \nonumber \\
&= \frac{\sum_{i=1}^n  \sigma^2(x_i) \{\sum_{p=0}^k \frac{1}{p!}(x_i-x_0)^p \lambda^{p} \}^2 K_h^2(x_i-x_0) }{ \{\sum_{i=1}^n \{\sum_{p=0}^k \frac{1}{p!}(x_i-x_0)^p \lambda^{p} \}^2K_h(x_i-x_0) \}^2}. \nonumber
\end{align}
Applying the first-degree Taylor expansion for $\sigma^2(x_i)$ in a sufficiently small neighborhood of  $x_0$, we have 
$$\sigma^2(x_i) \approx \sigma^2(x_0) + (x_i-x_0) \frac{d\sigma^2(x_0)}{dx} \approx \sigma^2(x_0)$$
since the derivative $\frac{d\sigma^2(x_0)}{dx}$ is bounded on $[a,b]$, by Assumption (IV).  
Then we can obtain 
\begin{align}
\mathrm{Var}(\widehat{g}_k (x_0)|x_1,...,x_n)
& \approx \frac{1}{n} \frac{\int_a^b  \sigma^2(x_0) \{1+\sum_{p=1}^k \frac{1}{p!}(z-x)^p \lambda^{p} \}^2 K_h^2(z-x)f(z)dz }{ \{\int_a^b \{1+\sum_{p=1}^k \frac{1}{p!}(z-x)^p \lambda^{p} \}^2K_h(z-x) f(z)dz\}^2} \nonumber \\
&=\frac{1}{nh} \frac{\sigma^2(x_0)\int  \{1+\sum_{p=1}^k \frac{1}{p!}(hw)^p \lambda^{p} \}^2 K^2(w)f(x+hw)dw }{ \{\int \{1+\sum_{p=1}^k \frac{1}{p!}(hw)^p \lambda^{p} \}^2K(w) f(x+hw)dw\}^2} \nonumber \\
&\approx \frac{\sigma^2(x_0)R(K)}{nhf(x_0)}+o_p\left(\frac{1}{nh}\right)
\end{align}

\subsection{Proof of Theorem 3}

Under Model (\ref{equ:exp}), we have $g(x_0)=g(0)e^{\lambda x}.$
Using the results in Theorems 1 and 2, along with the mean squared error (MSE) formula:
 $$\mathrm{MSE}(\hat\theta)=\mathrm{Var}(\hat\theta)+\mathrm{Bias}^2(\hat\theta, \theta), \quad \text{where} \quad \hat\theta \text{ is an estimator of} \quad \theta, $$ 
we derive the asymptotically optimal bandwidth $h_{o,k}$. 
For odd $k$, 
$$\mathrm{MSE}(\hat{g}_k(x_0)|x_1,...,x_n) \approx \frac{\sigma^2R(K)}{nhf(x_0)}+\frac{1}{((k+1)!)^2}\lambda^{2k+2}e^{2\lambda x}h^{2k+2}\mu_{k+1}^2g^2(0).$$
To find the optimal bandwidth, take the derivative of the MSE with respect to $h$ and set it to zero:
$$\frac{d(\mathrm{MSE})}{dh} \approx -\frac{\sigma^2R(K)}{nh^2f(x_0)}+\frac{1}{\{(k+1)!\}^2}\lambda^{2k+2}e^{2\lambda x}(2h+2)h^{2k+1}\mu_{k+1}^2g^2(0)=0.$$
Solving for $h$, the optimal bandwidth for DE1-$k$ becomes:
\begin{equation}
h_{o,k}^{2k+3}= \frac{\sigma^2R(K)((k+1)!)^2}{nf(x_0)e^{2\lambda x}\lambda^{2k+2}(2k+2)\mu_{k+1}^2g^2(0)}. \nonumber
\end{equation}
When the degree is $(k+2)$, the optimal bandwidth is given by
\begin{equation}
h_{o,k+2}^{2k+7}= \frac{\sigma^2R(K)((k+3)!)^2}{nf(x_0)e^{2\lambda x}\lambda^{2k+6}(2k+6)\mu_{k+3}^2g^2(0)}. \nonumber
\end{equation}
For a normal kernel $K_h$ and odd $k$, the moments are: $\mu_{k+1}$ = $k$!, and $\mu_{k+3}$ = $(k+2)$!.
The ratio of the successive optimal bandwidths is 
\begin{equation}
\frac{h_{o,k+2}^{2k+7}}{h_{o,k}^{2k+3}} = \frac{(k+3)^2(k+2)^2(2k+2)\mu_{k+1}^2}{(2k+6)\mu_{k+3}^2} = \frac{(k+3)(k+1)}{\lambda^4}.\nonumber
\end{equation}
Therefore, a recursive formula for the optimal bandwidth for odd $k$ is:
\begin{equation}
h_{o,k+2} =\Big(\frac{(k+3)(k+1)}{\lambda^4}h_{o,k}^{2k+3}\Big)^{1/(2k+7)}.  \nonumber
\end{equation}
Similarly, for even $k$, we obtain:
\begin{equation}
h_{o,k}^{2k+5}= \frac{\sigma^2R(K)((k+1)!)^2}{nf(x_0)e^{2\lambda x}\lambda^{2k+2}(2k+4)\mu_{k+2}^2(\lambda+\frac{f'(x_0)}{f(x_0)})^2}
\mbox{ and }
h_{o,k+2} =\Big(\frac{(k+2)^3}{(k+4)\lambda^4}h_{o,k}^{2k+5}\Big)^{1/(2k+9)}. \nonumber
\end{equation}

\section{Additional Simulation Results under Sparse Designs}

\subsection{More Designs for Example 2 (Section 5)}

\begin{figure}
        \centering
        \vspace{-10mm}
        \includegraphics[width=0.85\textwidth]{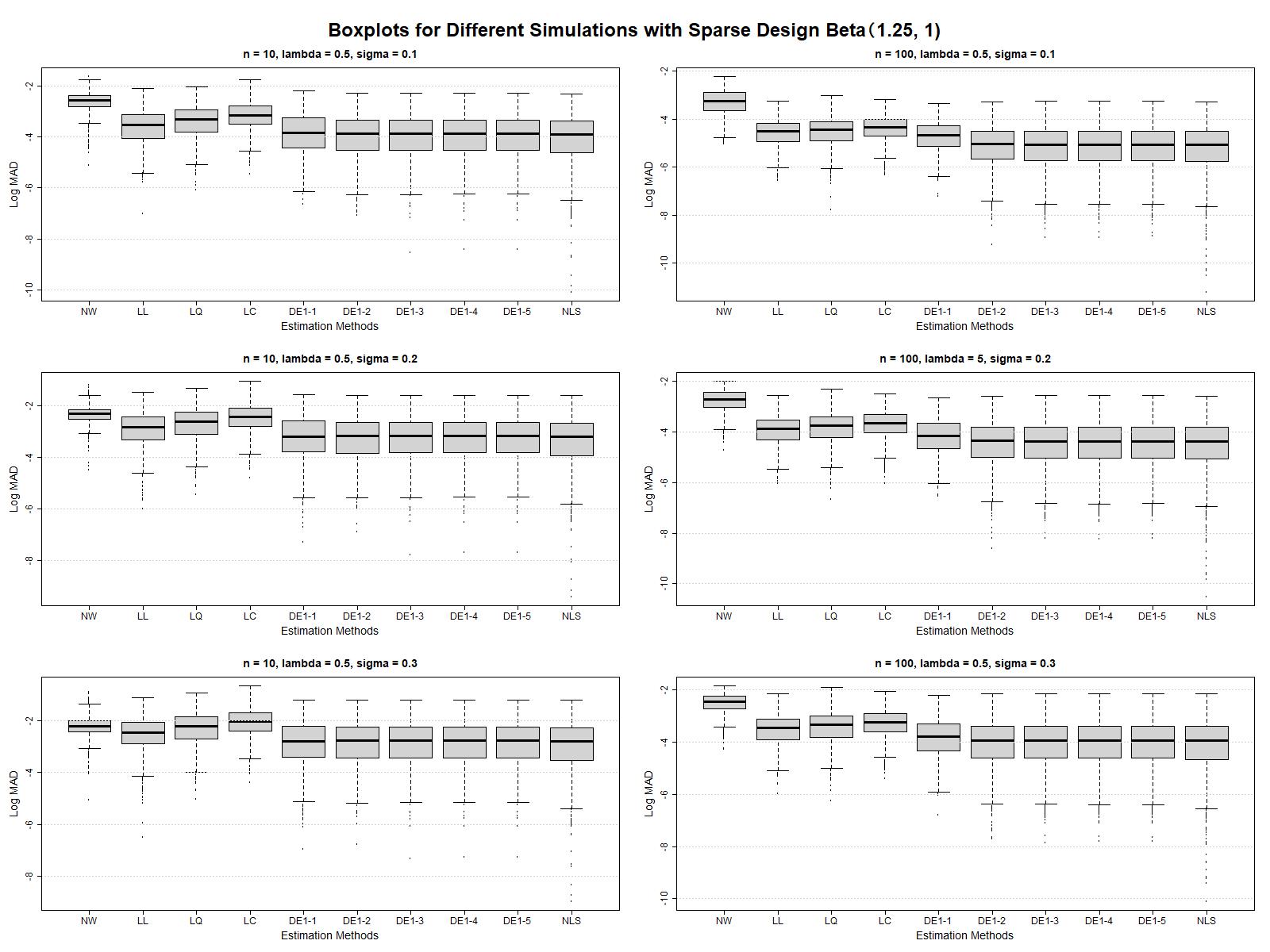}
        \vspace{-4mm}
        \caption{MAD (Median Absolute Deviation) distributions (on the log scale) for the estimation methods
applied to simulated data in sparse design Beta(1.25, 1). Sample size n = 30, 100. $\lambda$ = 0.5. }
        \label{fig:Beta1251}
        \end{figure}

\begin{figure}
        \centering
        \vspace{-8mm}
        \includegraphics[width=0.85\textwidth]{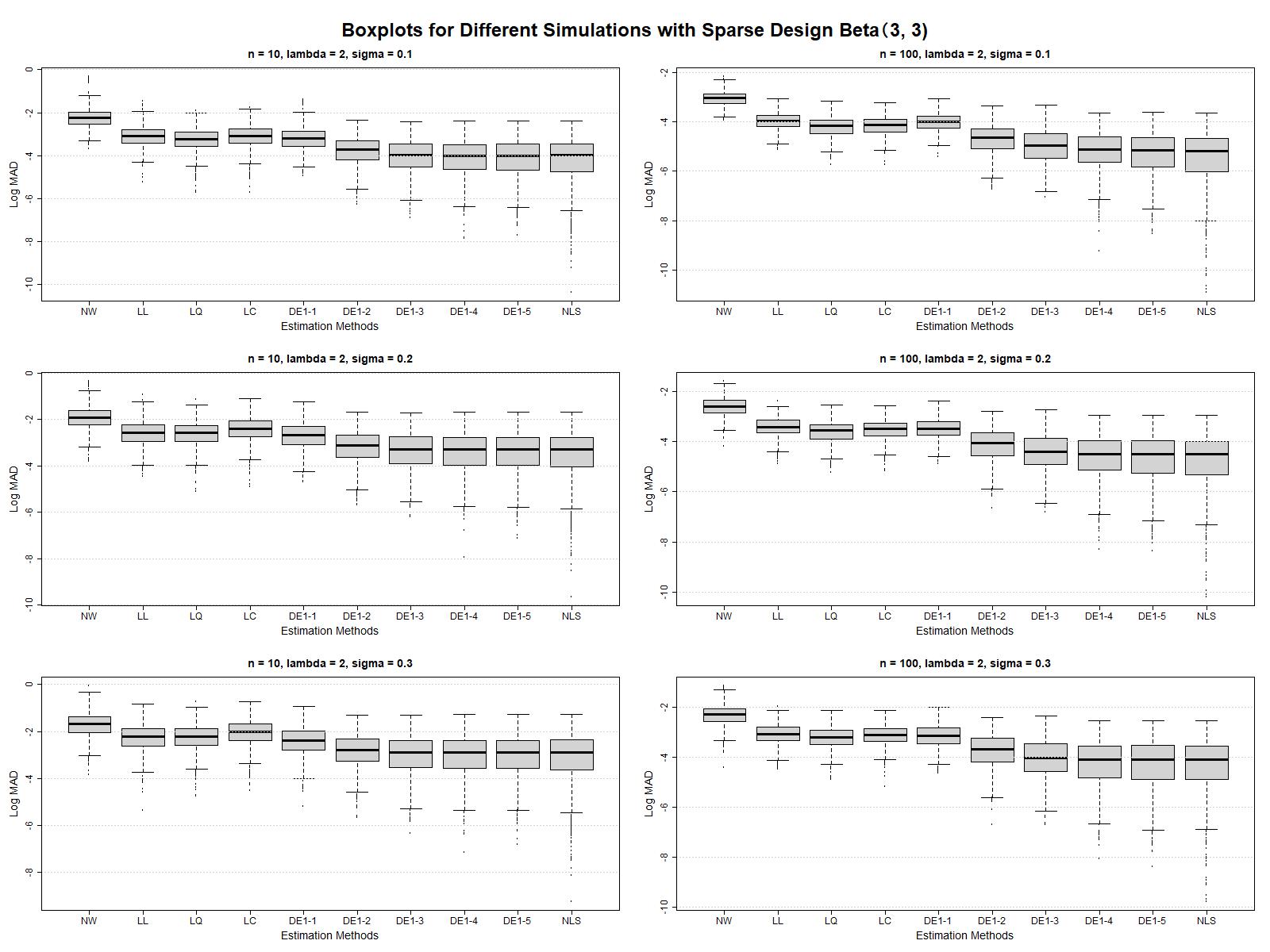}
        \vspace{-4mm}
        \caption{MAD (Median Absolute Deviation) distributions (on the log scale) for the estimation methods
applied to simulated data in sparse design Beta(3, 3). Sample size n = 10, 100. $\lambda$ = 2. }
        \label{fig:Beta33}
        \end{figure}

We explored additional sparse designs for normally distributed noise with $\sigma$ = 0.1, 0.2, 0.3 considering a Beta(1.25, 1) design with  $\lambda$ = 0.5 (Figure \ref{fig:Beta1251}) and  a  Beta(3, 3) design, with $\lambda$ = 2 (Figure \ref{fig:Beta33}). 

\begin{figure}
        \centering
        \vspace{-8mm}
        \includegraphics[width=0.85\textwidth]{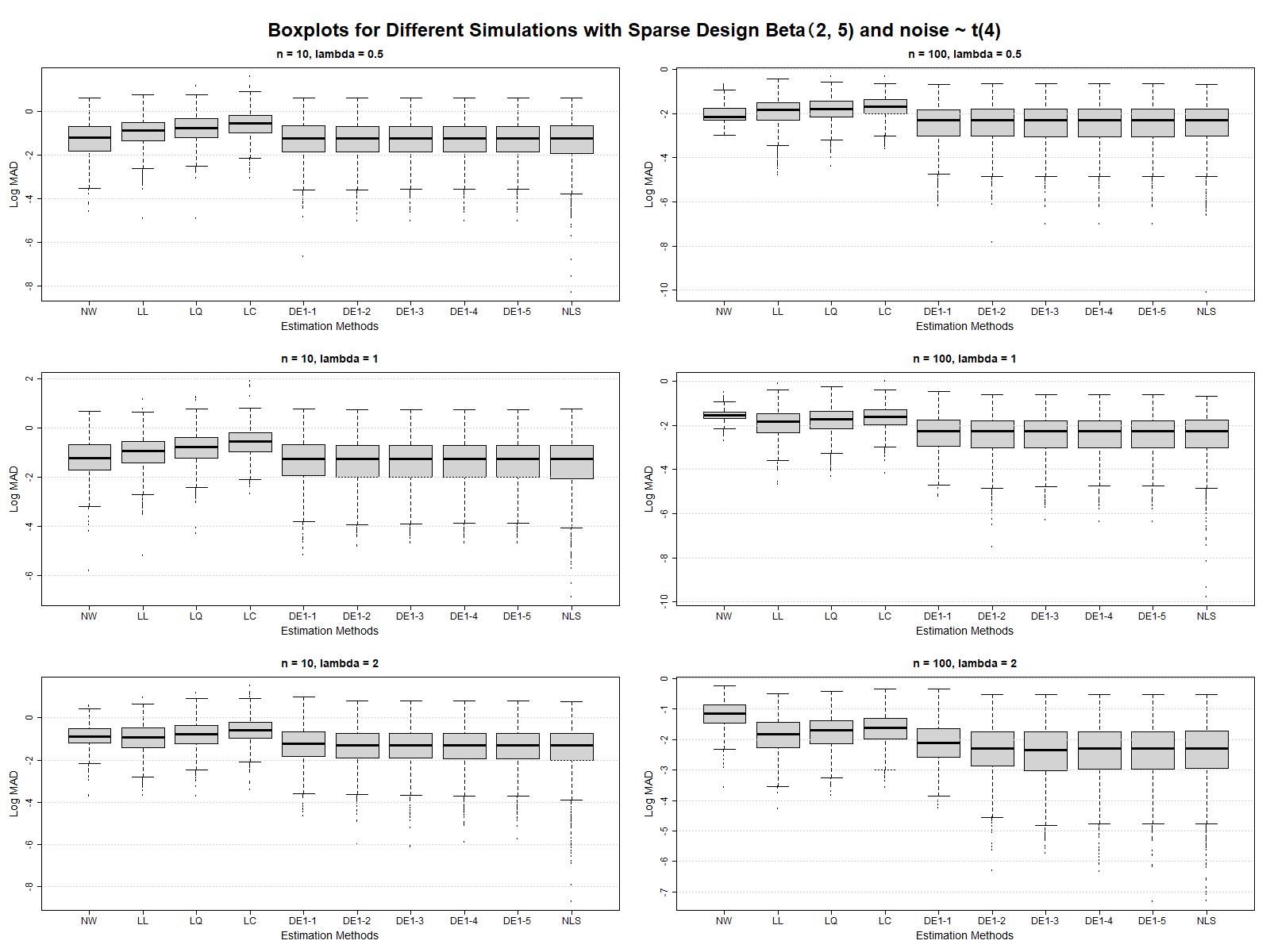}
        \vspace{-4mm}
        \caption{MAD (Median Absolute Deviation) distributions (on the log scale) for the estimation methods
applied to simulated data in sparse design Beta(2, 5). Noise $\sim$ t(4). Sample size n = 10, 100. }
        \label{fig:Beta25}
        \end{figure}

\begin{figure}
        \centering
        \vspace{-8mm}
        \includegraphics[width=0.85\textwidth]{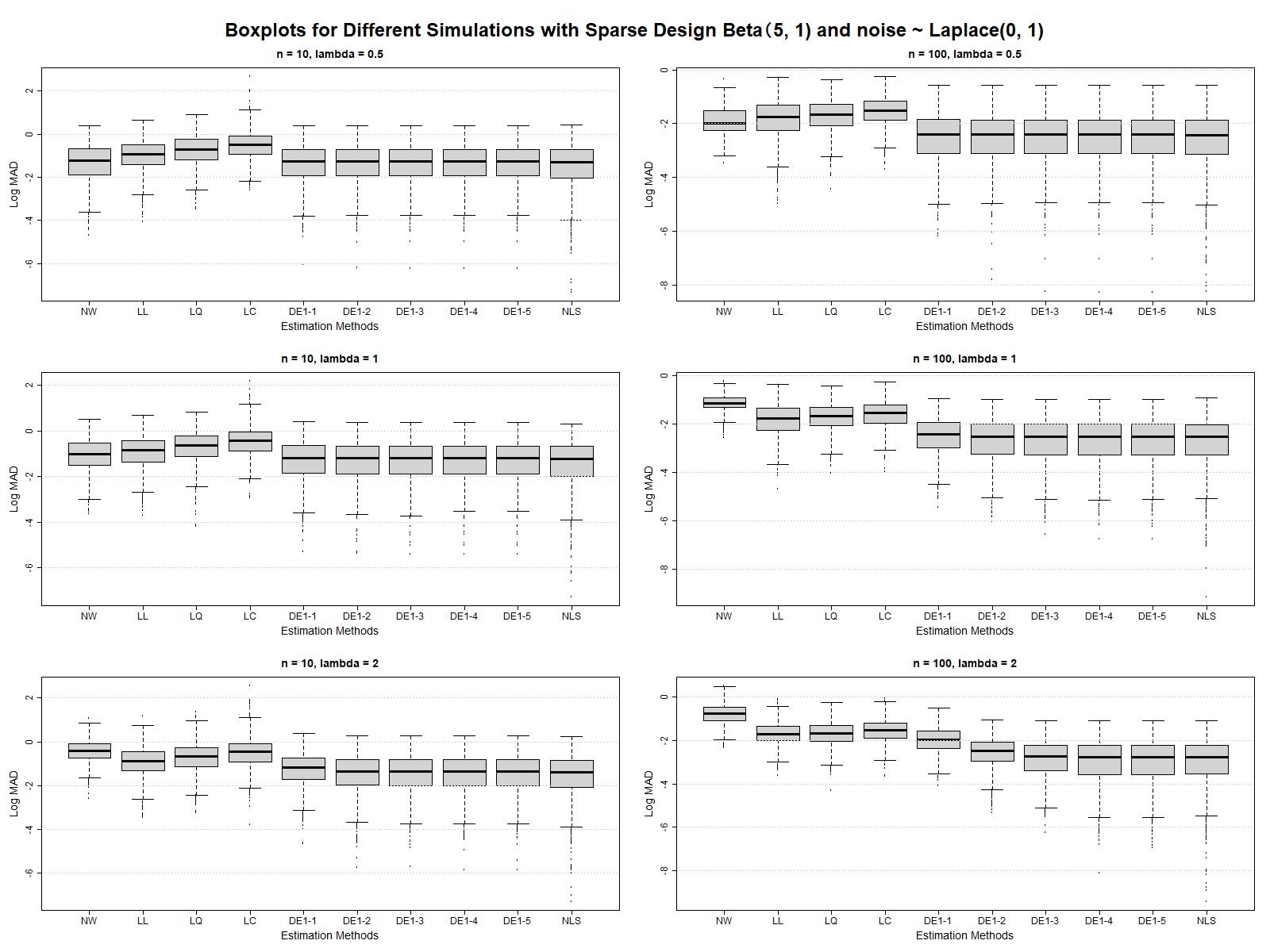}
        \vspace{-4mm}
        \caption{MAD (Median Absolute Deviation) distributions (on the log scale) for the estimation methods
applied to simulated data in sparse design Beta(5, 1). Noise $\sim$ Laplace(0, 1). Sample size n = 10, 100. }
        \label{fig:Beta51}
        \end{figure}

Additionally, we applied the DE-constrained approach to sparse designs with t-distributed noise ($\varepsilon \sim $ t(4), Figure \ref{fig:Beta25}) and Laplace-distributed noises ($\varepsilon \sim$ Laplace(0,1) (Figure \ref{fig:Beta51}).

The simulation results demonstrate the competitive performance of the DE-constrained approaches.

\subsection{More Designs for Example 3 (Section 5)}

\begin{figure}
        \centering
        \vspace{-4mm}
        \includegraphics[width=1\textwidth]{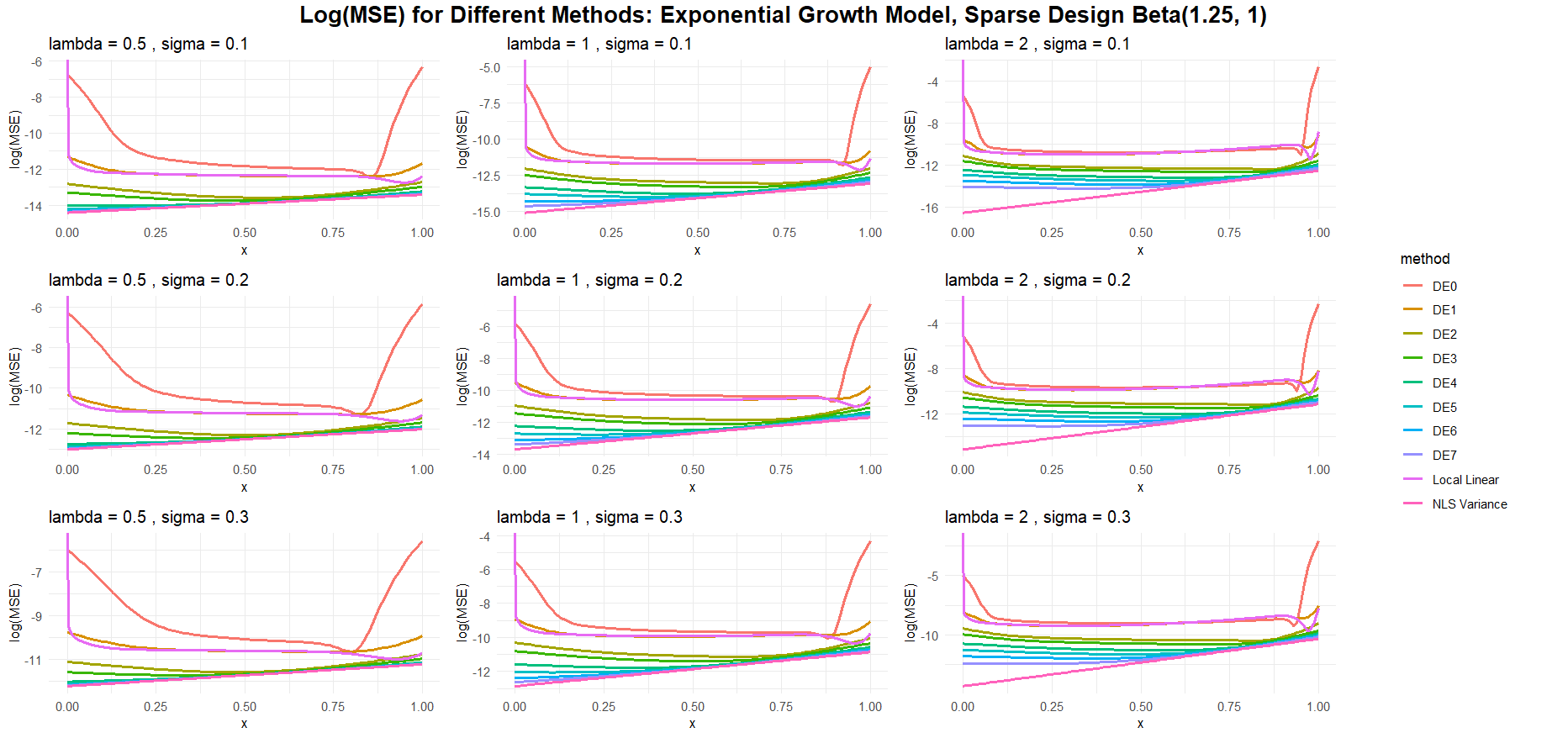}
        \vspace{-4mm}
        \caption{Log(MSE) for DE-constrained regression and local polynomial regression methods applied to simulated data with a sparse design Beta(1.25, 1) at different noise levels, $\sigma$ = 0.1, 0.2, 0.3. Sample size n = 10000. $\lambda$ = 0.5, 1, 2. }
        \label{fig:logMSEBeta1251}
        \end{figure}

\begin{figure}
        \centering
        \vspace{-4mm}
        \includegraphics[width=1\textwidth]{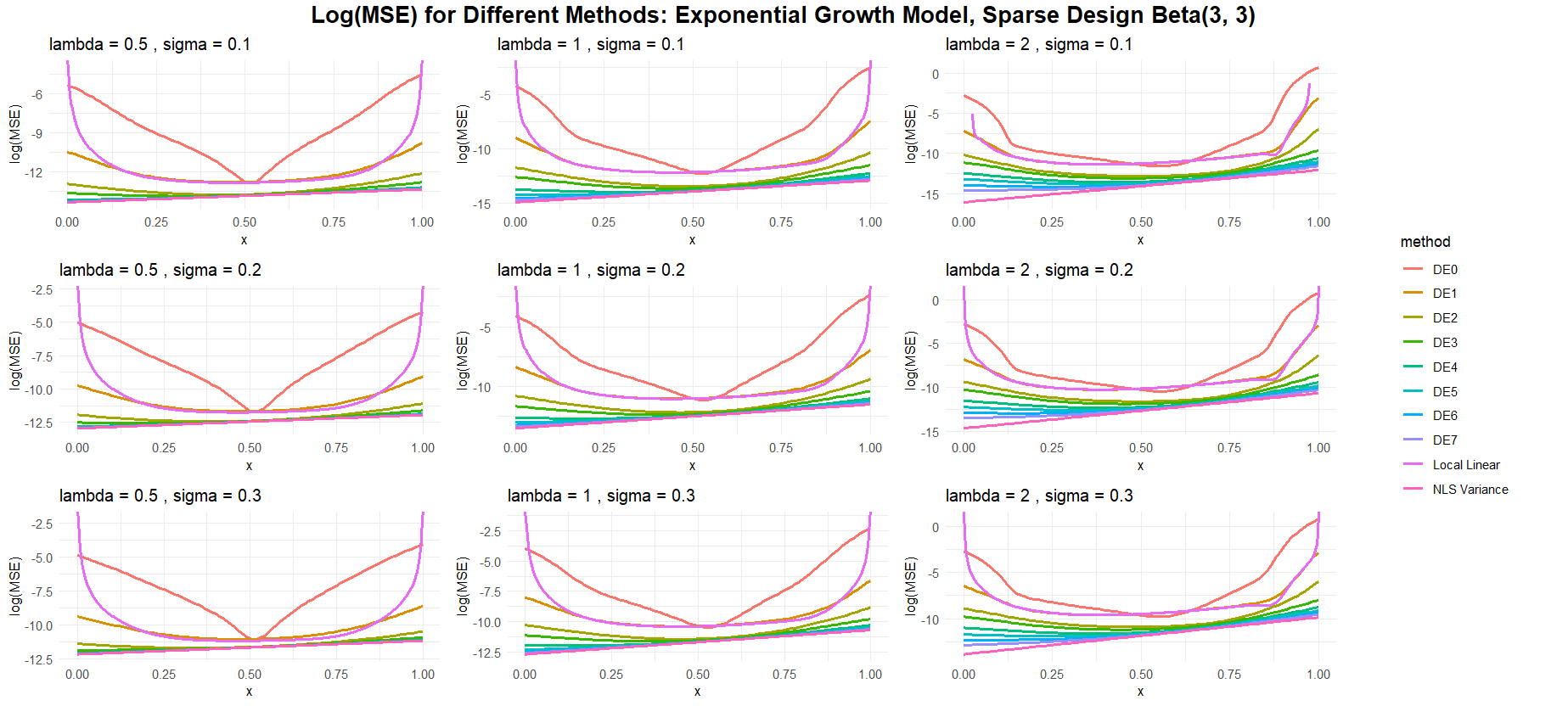}
        \vspace{-4mm}
        \caption{Log(MSE) for DE-constrained regression and local polynomial regression methods applied to simulated data with a sparse design Beta(3, 3) at different noise levels, $\sigma$ = 0.1, 0.2, 0.3. Sample size n = 10000. $\lambda$ = 0.5, 1, 2. }
        \label{fig:logMSEBeta33}
        \end{figure}

We explored additional sparse designs for normally distributed noise with $\sigma$ = 0.1, 0.2, 0.3 considering a Beta(1.25, 1) design (Figure \ref{fig:logMSEBeta1251}) and  a Beta(3, 3) design (Figure \ref{fig:logMSEBeta33}).  

\end{appendices}

\end{document}